\documentclass[11pt,twoside,a4paper,cmspaper,final,collab]{cms-tdr}
\def\svnVersion{cc7f996-D}\def\svnDate{2025/12/19}\def\cmsCernNoTag{CERN-EP-2024-331}\def\cmsCernDate{\today}\def\cmsMessage{Submitted to Physics Letters B}
\begin{document}\cmsNoteHeader{HIN-21-019}

\newcommand {\ptg}      {\ensuremath{\pt^{\PGg}}\xspace}
\newcommand {\see}      {\ensuremath{\sigma_{\eta\eta}}\xspace}
\newcommand {\ptj}      {\ensuremath{\pt^{\text{jet}}}\xspace}
\newcommand {\etag}     {\ensuremath{\eta^{\PGg}}\xspace}
\newcommand {\etaj}     {\ensuremath{\eta^{\text{jet}}}\xspace}
\newcommand {\phij}     {\ensuremath{\phi^{\text{jet}}}\xspace}
\newcommand {\hybrid}    {\ensuremath{\textsc{Hybrid}}\xspace}
\newcommand {\jewel}    {\ensuremath{\textsc{Jewel}}\xspace}
\newcommand {\pyquen}   {\ensuremath{\textsc{Pyquen}}\xspace}
\newcommand {\xjg}      {\ensuremath{x_{{\mathrm{j}}\PGg}}\xspace}
\newcommand {\dphijg}   {\ensuremath{\Delta\phi_{{\mathrm{j}}\PGg}}\xspace}
\newcommand {\ddr}      {\ensuremath{\Delta{j}}\xspace}
\newcommand {\PbPb}     {\ensuremath{\mathrm{PbPb}}\xspace}
\newcommand {\pp}       {\ensuremath{\Pp\Pp}\xspace}
\newcommand {\Escheme}  {\ensuremath{E\text{--scheme}}\xspace}
\newcommand {\Sumiso}   {\ensuremath{I_{\PGg}^\text{Reco}}\xspace}
\newcommand {\Sumisogen}   {\ensuremath{I_{\PGg}^\text{Gen}}\xspace}
\newlength              \cmsTabSkip\setlength{\cmsTabSkip}{1ex}

\cmsNoteHeader{HIN-21-019}

\title{First measurement of jet axis decorrelation with photon-tagged jets in \texorpdfstring{\pp}{pp} and \texorpdfstring{\PbPb}{PbPb} collisions at \texorpdfstring{5.02\TeV}{5.02 TeV}}

\date{\today}

\abstract{
The first measurement of the jet axis decorrelation in events with jets recoiling from an isolated photon is presented for lead-lead (\PbPb) and proton-proton (\pp) collisions at a nucleon-nucleon center-of-mass energy of 5.02\TeV. The jet axis decorrelation is the angular difference ($\Delta{j}$) between two definitions of the jet axis. This quantity is expected to be sensitive to the scattering of jet constituents in the quark-gluon plasma (QGP). Events which have a leading isolated photon with transverse momentum $60 < \ptg < 200\GeV$ and recoiling jets with $30 <\ptj < 100\GeV$ are selected for the analysis. The \PbPb result is reported as a function of collision centrality and compared to \pp reference data. Jets with $\ptj < 60\GeV$ have consistent $\Delta{j}$ shapes for \pp and \PbPb collisions. However, a narrowing is observed for jets with $\ptj > 60\GeV$ in central \PbPb collisions. The results are compared to predictions from the \jewel, \pyquen and \hybrid theoretical models, which include different descriptions of parton energy loss in the QGP.
}

\hypersetup{
pdfauthor={CMS Collaboration},
pdftitle={First measurement of the jet axis decorrelation with photon-tagged jets in pp and PbPb collisions at 5.02 TeV with CMS},
pdfsubject={CMS},
pdfkeywords={CMS, heavy ion}}

\maketitle

\section{Introduction}
\label{sec:introduction}

The quark-gluon plasma (QGP) is created when two ultra-relativistic heavy ions collide resulting in a deconfined state of quarks and gluons that exhibits liquid-like flow~\cite{Collins:1974ky, Karsch:1995sy, Busza:2018rrf}. The QGP medium is formed on a time scale of around $1\unit{fm}/c$~\cite{Bjorken:1982qr}. In the initial collision, pairs of partons from the colliding ions occasionally undergo a high-energy scattering with large momentum transfer. These scatterings occur before the QGP formation, so their production process is well-understood and unaffected by the QGP. As the hard-scattered partons pass through the QGP, they interact with it via elastic scattering from medium particles and medium-induced gluon radiation, resulting in energy loss and broadening of the momentum distribution transverse to the jet axis~\cite{Appel:1985dq, Blaizot:1986ma, Gyulassy:1990ye, Wang:1991xy, Baier:1996sk, Zakharov:1997uu}. This phenomenon is known as ``jet quenching'' and can be used as a probe of QGP properties~\cite{Connors:2017ptx, Cunqueiro:2021wls, Apolinario:2022vzg}. A review of QGP studies with the CMS experiment can be found in Reference~\cite{CMS:2024krd}.

Measurements of hadrons and jets with high transverse momentum (\pt) formed the basis of initial jet quenching studies~\cite{Adams:2003kv, Adare:2008qa, Abelev:2012hxa, Aad:2015wga, CMS:2012aa, Khachatryan:2016odn, Chatrchyan:2011sx, Aad:2010bu, ATLAS:2014cpa, Adam:2015doa, Khachatryan:2016jfl, Adam:2015ewa, Adamczyk:2016fqm}. Jet substructure measurements further clarify the picture of jet energy loss; observables can be designed to be sensitive to different energy loss processes, such as medium-induced wide-angle gluon radiation, Moli\`ere scattering of jet constituents from medium particles, and wake effects arising from medium particles perturbed along the jet's direction of motion~\cite{hulcher2022sensitivity}.

This Letter presents a measurement of the jet axis decorrelation, which is a novel tool for investigating transverse momentum broadening and medium-induced radiation within the QGP. The jet axis decorrelation, \ddr, is defined as the angular difference between two definitions of the jet axis: the energy-weight (\Escheme) axis and the winner-take-all (WTA) axis,
\begin{equation} \label{eq:dj}
    \ddr = \sqrt{\left(\etaj_{\Escheme} - \etaj_{\mathrm{WTA}}\right)^{2} + \left(\phij_{\Escheme} - \phij_{\mathrm{WTA}}\right)^{2}},
\end{equation}
where $\etaj_{\Escheme}$ and $\etaj_{\mathrm{WTA}}$ are the pseudorapidities of the axes, and $\phij_{\Escheme}$ and $\phij_{\mathrm{WTA}}$ are their azimuthal angles~\cite{Cal:2020}. The \Escheme axis is the standard anti-\kt jet axis, which is determined by summing the four-momenta of the two subjets at each step of the jet clustering algorithm~\cite{Cacciari:2008gp}. In contrast, the WTA axis is found by reclustering the jet with the Cambridge--Aachen algorithm and taking the direction of the hardest subjet at each step as the axis~\cite{Bertolini:2013iqa}. The resulting \Escheme and WTA axes align with the direction of the average and leading energy flow in the jet, respectively. These axes are expected to decorrelate even in vacuum because the parton showering process alters them from the original parton's direction. In heavy ion collisions, interactions with quark-gluon plasma constituents could cause the jet axes to further decorrelate from each other. The WTA axis follows only the hardest jet subcomponent, so it has reduced sensitivity to soft radiation and medium response and may be more sensitive to Moli\`ere scattering. In contrast, the \Escheme axis may be less modified by the medium due to conservation of energy. These characteristics make measurements of \ddr useful for understanding medium scattering effects.

However, all inclusive jet and dijet measurements suffer from a ``survivor bias,'' since jets which are wider, have more constituents, or travel farther through the QGP may be more strongly quenched and, therefore, fall below the selection of a minimum jet \pt. In combination with the steeply falling jet \pt spectrum, this can bias measurements in a set \ptj interval towards narrower and less quenched jets~\cite{Brewer:2021hmh}. The jet axis decorrelation was previously measured in inclusive jets by the ALICE experiment~\cite{ALICE:2023dwg}. This analysis found that the \ddr distribution peaks at smaller values of \ddr in \PbPb collisions than \pp collisions, which is consistent with survivor bias effects.

To mitigate bias, substructure measurements can instead study jets tagged by electroweak bosons such as photons or \PZ bosons~\cite{Kartvelishvili:1995fr, Wang:1996yh, Wang:1996pe, Dai:2012am,Kang:2017xnc}. Electroweak bosons are colorless and do not interact via the strong force, so on average their transverse momentum reflects the initial energy of the associated high-energy scattered partons before they lose energy in the QGP and fragment into jets~\cite{Aad:2015lcb, Chatrchyan:2012vq, Chatrchyan:2012nt, Chatrchyan:2014csa, CMS:2020oen}. Furthermore, the production of high-\pt jets recoiling from electroweak bosons is dominated by quark fragmentation and is, therefore, theoretically better-controlled compared to inclusive jet production, which includes a mix of quark and gluon jets~\cite{Neufeld:2010fj}.

In this Letter, the selected electroweak bosons are isolated photons, which have higher background but also significantly higher production rate than \PZ bosons. The photon isolation requirement suppresses background from photons that originate from neutral meson decays and yields a sample with predominantly prompt photons. Prompt photons arise from an initial high-energy scattering or the subsequent fragmentation which occurs before hadronization~\cite{Aad:2015lcb, Chatrchyan:2012vq}. Selecting jets with an isolated photon tag allows a less-biased comparison between jets in \pp and \PbPb collisions. Such photon-tagged jet studies also facilitate the extraction of QGP properties via comparisons with theoretical models~\cite{Wang:2013cia, Casalderrey-Solana:2014bpa, Casalderrey-Solana:2015vaa, KunnawalkamElayavalli:2016ttl, CMS:2012ytf, CMS:2021otx, Sirunyan:2017qhf}.

This measurement of the photon-tagged jet axis decorrelation complements other studies of photon-tagged jet substructure, such as the photon-tagged jet shape and  fragmentation function, which are primarily sensitive to hadronization effects and the medium response, but insensitive to Moli\`ere scattering~\cite{Sirunyan:2018ncy, Sirunyan:2018qec, ATLAS:2019dsv}. The analysis also builds on studies of the photon-tagged groomed jet radius and girth~\cite{CMS:2024zjn} by investigating jet survivor bias with a new observable. Finally, the comparison between \PbPb collisions and \pp collisions provides information about parton shower modifications due to interactions with the QGP. Tabulated results are provided in the \textsc{HepData} record for this analysis~\cite{hepdata}.

\section{Experimental setup}
\label{sec:detector}

The central feature of the CMS apparatus is a superconducting solenoid of 6\unit{m} internal diameter, providing a magnetic field of 3.8\unit{T}. Within the solenoid volume are a silicon pixel and strip tracker, a lead tungstate crystal electromagnetic calorimeter (ECAL), and a brass and scintillator hadron calorimeter (HCAL), each composed of a barrel and two endcap sections. Forward hadronic calorimeters (HF) extend the pseudorapidity coverage provided by the barrel and endcap detectors. Muons are measured in gas-ionization detectors embedded in the steel flux-return yoke outside the solenoid. A more detailed description of the CMS detector, together with a definition of the coordinate system used and the relevant kinematic variables, can be found in Ref.~\cite{CMS:2008xjf}. Reference \cite{Hayrapetyan_2024} summarizes the CMS detector developments since Run 1.

\section{Event selection}

The \pp and \PbPb data samples were collected by the CMS experiment at a nucleon-nucleon center-of-mass energy of $\sqrtsNN = 5.02\TeV$ in 2017 and 2018, respectively. The \PbPb sample corresponds to an integrated luminosity of $1.69 \pm 0.03\nbinv$, while the \pp sample corresponds to $302 \pm 6\pbinv$~\cite{CMS-LUM-17-003, CMS:2025rzq, CMS-PAS-LUM-19-001}.

Events of interest are selected online using a two-tiered trigger system~\cite{Khachatryan:2016bia}. The first level (L1), composed of custom hardware processors, uses information from the ECAL to select events containing a transverse energy (\et) deposit greater than 21\GeV. The second level, known as the high-level trigger (HLT), consists of a farm of processors running a version of the full event reconstruction software optimized for fast processing. At this stage, events are required to contain at least one photon candidate with transverse momentum $\ptg > 40\GeV$.

Both \pp and \PbPb events are required to have at least one reconstructed primary interaction vertex less than 15\cm from the center of the interaction region along the beam direction. In addition, in order to select inelastic \PbPb collisions, these events are required to have at least two calorimeter towers in the HF on each side of the interaction point with energy greater than 4\GeV. To further reject beam-gas interactions, the lengths of pixel clusters along the beam direction are also required to be compatible with particles originating from the primary vertex~\cite{Khachatryan:2010us}.

Nuclei are extended objects, and their collisions can be characterized by their degree of overlap, or centrality, which is expressed as a percentile of the total inelastic hadronic cross section. The most central collisions, \ie those with the smallest impact parameters, are denoted with a centrality close to 0\%. The collision centrality can be estimated from the sum of the transverse energy of all hadronic collision products in the HF calorimeter~\cite{Chatrchyan:2011sx}. After correcting for event selection efficiency and accounting for contamination from electromagnetic processes, the distribution of the total transverse energy is divided into equal partitions and used to classify events into centrality intervals. The analysis is restricted to events with centrality in the 0--90\% range, where the event selection is fully efficient.

\section{Event reconstruction}

The global event reconstruction (also called particle-flow event reconstruction~\cite{Sirunyan:2017ulk}) aims to reconstruct and identify each individual particle in an event, with an optimized combination of all subdetector information. In this process, the identification of the particle type (photon, electron, muon, charged hadron, neutral hadron) plays an important role in the determination of the particle direction and energy. Photons are identified as ECAL energy clusters not linked to the extrapolation of any charged particle trajectory to the ECAL. Electrons are identified as a primary charged particle track and potentially many ECAL energy clusters corresponding to this track extrapolation to the ECAL and to possible bremsstrahlung photons emitted along the way through the tracker material. Charged hadrons are identified as charged particle tracks neither identified as electrons, nor as muons. Finally, neutral hadrons are identified as HCAL energy clusters not linked to any charged hadron trajectory, or as a combined ECAL and HCAL energy excess with respect to the expected charged hadron energy deposit.

The energy of charged hadrons is determined from a combination of the track momentum and the corresponding ECAL and HCAL energies, corrected for the response function of the calorimeters to hadronic showers. The energy of neutral hadrons is obtained from the corresponding corrected ECAL and HCAL energies. The energy of electrons is determined from a combination of the track momentum at the main interaction vertex, the corresponding ECAL cluster energy, and the energy sum of all bremsstrahlung photons attached to the track. Finally, the energy of photons is obtained from the ECAL measurement using the global event description algorithm~\cite{CMS:2020uim}. In \PbPb collisions, the photon energy scale is corrected with an additional multivariate regression analysis using simulated events that contain high-\pt photons. No additional correction is applied in \pp collisions, since the nominal photon energy scale is already sufficient.

For each event, hadronic jets are clustered from these reconstructed particles using the infrared- and collinear-safe anti-\kt algorithm~\cite{Cacciari:2008gp, Cacciari:2011ma} with a distance parameter of $R = 0.3$, which is chosen to reduce the effect of underlying-event (UE) fluctuations. To correct for the UE addition to reconstructed \ptj in \PbPb collisions, a particle-level pileup subtraction technique, also known as ``constituent subtraction,'' is used~\cite{Berta:2014eza}. This algorithm removes or corrects jet constituents for the uncorrelated background on an event-by-event basis using the average UE density in a given $\eta$ region.

Jet energy corrections are derived from simulation to bring the measured response of jets to that of particle level jets on average. In situ measurements of the momentum balance in dijet, \GAMJET, $\PZ+\text{jet}$, and multijet events are used to account for any residual differences in the jet energy scale between data and simulation~\cite{Khachatryan:2016kdb}. Additional selection criteria are applied to each jet to remove jets potentially dominated by anomalous contributions from various subdetector components or reconstruction failures~\cite{CMS:2017wyc}. In \PbPb collisions, an additional $\phi$-dependent jet energy correction was derived to deal with effects from the interplay between the UE subtraction and azimuthal asymmetries in the data, including those due to detector nonuniformity.

\section{Analysis method}
\label{sec:analysis}

The analysis considers \GAMJET pairs formed between the highest \pt isolated photon candidate and all jets in the same event with azimuthal angle difference $\abs{\dphijg} > 2\pi/3$. Photon candidates are required to have transverse momentum $60 < \ptg < 200\GeV$ and to be located in the barrel region of the ECAL ($\abs{\etag} < 1.44$), which has the best performance for photon reconstruction. Reconstructed jets are required to have $30 < \ptj < 200\GeV$ and to be situated within $\abs{\etaj} < 1.6$, where the jet energy resolution (JER) is well understood. Typically, each photon is paired with at most one jet.

Isolated photons are defined as those where the \et sum of generator-level particles within $\Delta R = \sqrt{\smash[b]{(\Delta\eta)^2+(\Delta\phi)^2}} = 0.4$ around the photon, \Sumisogen, is less than 5\GeV. To select isolated photons at reconstruction-level, a requirement is imposed on the energy surrounding the photon, \Sumiso, which is defined as the \et sum of final-state reconstructed particles in a cone of radius $\Delta R = 0.3$ around the photon cluster center, not including the photon cluster energy. In \PbPb collisions, the estimated UE energy is subtracted from \Sumiso to account for its variation with collision centrality.

To optimize selection of reconstructed photons corresponding to the generator-level definition, additional requirements are imposed on the reconstructed photon's electromagnetic shower shape and hadronic energy fraction. These selections are optimized using binary decision trees~\cite{CMS:2020uim}. Photons with large \Sumiso, wide shower shapes, or high hadronic energy fractions are dominated by background. Finally, photons that are matched to an electron candidate are also discarded~\cite{Chatrchyan:2012vq}. The L1+HLT trigger selection is 98--100\% efficient for isolated photons with $\ptg > 60\GeV$. The remaining inefficiency is corrected independently in \pp collisions and each \PbPb centrality interval.

In \PbPb collisions, there can be uncorrelated jets arising either from UE fluctuations misidentified as true hard-scattering products or from unrelated binary nucleon-nucleon collisions in the same event. The uncorrelated jet contribution to the \ddr distribution must be subtracted in order to isolate the effects of medium interaction with jets produced in the same hard scattering as the photon. This background contribution is estimated by taking each selected photon and reconstructing the observable with jets from independent \PbPb data events, collected from minimum-bias triggers, which have similar centrality; the sum of transverse momentum in the HF calorimeter in matched events is required to be within 10\% of the photon event. This separately-triggered event set is randomly sampled 100 times per photon event and used to construct the average uncorrelated jet background contribution to the \ddr distribution. This background contribution is then subtracted from the in-event \GAMJET \ddr distribution, and the final \ddr distribution is normalized per photon.

After the uncorrelated background jet contribution is subtracted, there is still a contribution from jets paired with nonprompt photons that pass the reconstruction-level isolation requirements. Nonprompt photons predominantly arise from neutral meson decays instead of coming from an initial high-energy scattering. Their contribution can be removed with a technique called purity subtraction, which uses the electromagnetic shower shape variable, \see, and the reconstruction-level isolation sum, \Sumiso~\cite{Khachatryan:2010fm, AWES1992130, Sirunyan:2017qhf}. Photons with a large \Sumiso or large \see are almost entirely nonprompt photons.

First, a template fit of the electromagnetic shower shape \see is used to determine the photon purity, which is the fraction of prompt photons in each \ptg interval. The signal templates are obtained from \PYTHIA~\cite{Sjostrand:2014zea} 8.230 Monte Carlo simulation using the CP5 tune~\cite{CMS:2019csb} to model isolated prompt \GAMJET processes; in \PbPb collisions this signal is additionally embedded in \HYDJET 1.9 simulation to model the UE~\cite{Lokhtin:2008xi}. The background templates are obtained from the nonisolated sideband region in data, where $10 < \Sumiso < 20\GeV$. This region is dominated by decay photons and provides a good description of the background \see shape. Next, the \ddr distribution from pairs of nonprompt photons and jets is evaluated using events with a large photon shower width, from $0.012 < \see < 0.02$, which is also dominated by nonprompt photons. The background contribution is then subtracted from the yield for the signal events, which have $\see < 0.01$, according to the purity values obtained from the template fits.

Finally, the measurement is unfolded in order to deconvolve detector smearing of the \ptj and \ddr distributions and allow comparison with theoretical predictions. The data are unfolded simultaneously in bins of \ptj and \ddr, with separate response matrices for \pp collisions and for each \PbPb centrality interval. The response matrices are constructed using \PYTHIA 8.230 CP5 tune Monte Carlo simulation for \pp collisions and \PYTHIA 8.230 CP5 tune + \HYDJET 1.9 Monte Carlo simulation for \PbPb collisions. For \PbPb collisions, the average UE activity $\rho$ within the analysis acceptance $\abs{\etag} < 1.44$ in the simulation is matched to the distribution in data in each centrality interval~\cite{CMS:2021vui}. Simulated jet energies are smeared with the data/simulation JER difference, using scale factors obtained from $\sqrt{s} = 13\TeV$ \pp data taken in 2017 and 2018~\cite{Khachatryan:2016kdb}. Furthermore, the events used to construct the response matrices are weighted so that their photon \pt spectra matches that of data in the range $60 < \ptg < 200\GeV$.

The unfolding is performed with the D'Agostini iterative method with early stopping~\cite{DAgostini:1994fjx}, where the number of iterations serves as the regularization parameter. The prior used for the unfolding is taken from the generator-level response matrix. Two unfolding corrections are performed: a purity correction before unfolding to account for detector-level jets that are not matched to any generator-level jet, and an efficiency correction after unfolding to account for ``misses,'' where the generator-level jet is not reconstructed.

The unfolding regularization is selected by minimizing the mean squared error (MSE) of the unfolded result using pseudo-experiments~\cite{brenner2020comparison}. Theory inputs, unrelated to the \PYTHIA Monte Carlo simulation used to construct the response matrices, are forward folded with each response matrix and used to generate 1000 pseudo-experiments, which are unfolded at each iteration parameter and used to calculate the bias and the variance. The iteration with the minimum MSE is used to regularize data. Typical values are 2--10 iterations for \PbPb and 10--25 iterations for \pp. This method ensures that the bias (which increases with stronger regularization) and the variance (which decreases with stronger regularization) are balanced.

\section{Systematic uncertainties}

Systematic uncertainties are estimated independently for \pp data and for each \PbPb centrality interval and depend on the jet \pt interval and \ddr bin. The absolute contributions from each source, as well as the total systematic uncertainties, are summarized in Table~\ref{tab:systematics1} for $30 < \ptj < 60\GeV$ and in Table~\ref{tab:systematics2} for $60 < \ptj < 100\GeV$.

\emph{Electron rejection:} Despite applying the electron rejection criteria to all selected photons, some electrons may still contaminate the sample due to the inefficiencies in identifying electrons. To determine the corresponding uncertainty in the \ddr spectra, the result is calculated without applying any electron rejection criteria. The inefficiency of the electron selection measured with data (14\%) is then used to scale the variation in the final observable. This scaled variation is quoted as the systematic uncertainty due to remaining electrons.

\emph{Photon purity:} The photon purity estimation has an associated uncertainty arising from the background template choice; variations of the signal template yield a negligible change to the result. The nominal background template is obtained from a nonisolated sideband region, where $10 < \Sumiso < 20\GeV$. To calculate the systematic uncertainty, the boundaries of the background template \Sumiso requirement are shifted up and down by 5\GeV, and the purity fits are performed again with these alternate background templates~\cite{Sirunyan:2017qhf}. The extracted purity values are used to perform the purity subtraction and propagated through the analysis chain. The maximum variation in the final observable is quoted as the systematic uncertainty.

\emph{Isolation definition:} The difference between the photon reconstruction- and generator-level isolation definitions (described in Sec.~\ref{sec:analysis}) also results in a systematic uncertainty, since the finite detector resolution smears their correspondence. The impact of the photon isolation criteria smearing is studied with simulation by comparing the fully corrected and unfolded \ddr spectra with reconstruction- and generator-level photon isolation requirements. The proportional variation found in simulation is then scaled to data and quoted as the systematic uncertainty.

\emph{Photon energy scale:} The uncertainty in the photon energy scale is estimated by reconstructing the \PZ boson mass peak in simulation and data using electrons reconstructed as photon candidates. The \PZ boson mass peak is fit with a double-sided Crystal Ball function~\cite{Skwarnicki:1986xj}, and the difference in the means between data and simulation after the photon energy correction is used to determine the photon energy scale uncertainty. The resulting systematic uncertainty in the \ddr spectra is calculated by scaling the \ptg of each photon by the simulation/data difference and finding the difference in the final \ddr spectra.

\emph{Mixed-event background subtraction:} Mixed-event background subtraction begins to break down at low \ptj due to the very high jet occupancy in central \PbPb collisions. As a result, there is a residual nonclosure (a discrepancy between the reconstructed and true jet distributions) for $\ptj < 50\GeV$ and large nonclosure for $\ptj < 30\GeV$. To account for the effects of this nonclosure, the analysis is limited to jets with $\ptj > 30\GeV$, and a systematic uncertainty is assigned for remaining nonclosure, which is determined from simulation. Simulated data are analyzed with the nominal mixed-event background subtraction method and without it; in the latter case, reconstruction-level jets are matched with generator-level jets to determine the signal \ddr distribution. Finally, both \ddr distributions are purity-subtracted and unfolded. The percent nonclosure found in simulation is scaled to data and quoted as the systematic uncertainty. This is a dominant systematic at low \ptj in central \PbPb collisions.

\emph{Jet energy correction (JEC):} After applying all jet energy corrections, any remaining nonclosure in the JEC is corrected during the unfolding step. The uncertainty in this correction is calculated by altering the response matrices; when constructing the response matrices from simulation, the reconstructed jet energies are scaled up and down by the uncertainty in the JEC, which is found using dijet and \GAMJET balancing methods~\cite{Khachatryan:2016kdb}. The maximum difference in the unfolded result is quoted as the systematic uncertainty. This is a dominant systematic uncertainty in both \pp and \PbPb collisions.

\emph{Jet energy resolution (JER):} When constructing the response matrices, jets are smeared to correct for the data/simulation JER difference, using scale factors obtained from $\sqrt{s} = 13\TeV$ \pp data taken in 2017 and 2018~\cite{Khachatryan:2016kdb}, referred to as ``neighbor \pp.'' The uncertainty in this JER difference is added to the nominal difference to create new response matrices for unfolding the data. The resulting difference in the unfolded \ddr spectrum is quoted as the systematic uncertainty.

\emph{Centrality:} In \PbPb data, the event centrality is determined by the sum of the transverse energy in the HF calorimeters, which has an associated uncertainty. To determine the effect on the final result, the HF transverse energy sum boundaries for each centrality interval are varied up and down according to their uncertainties. The maximum resulting difference in the \ddr spectra in each centrality interval is taken as the systematic uncertainty.

\emph{Unfolding prior:} There are uncertainties associated with unfolding due to the prior choice, which can introduce a bias in the final result. Nominally, the prior used for unfolding is taken from the generator-level simulation used to construct the response matrix, which is \PYTHIA CP5 tune Monte Carlo for \pp collisions and \PYTHIA CP5 tune + \HYDJET Monte Carlo for \PbPb collisions. To determine the uncertainty, unfolding is performed with an alternate theoretical input as the prior. The \jewel \PbPb 0--10\% prediction is used as the alternate prior for unfolding \pp and \PbPb collisions, in order to contrast with the nominal \PYTHIA-based prior. The resulting variation in the unfolded \ddr spectra is taken as the systematic uncertainty. This systematic uncertainty is of moderate size in both \pp and \PbPb collisions.

\begin{table*}[hbt!]
    \centering
    \topcaption{Average of the bin-by-bin \ddr absolute, symmetric systematic uncertainties for jets with $30 < \ptj < 60\GeV$ in \pp data as well as in \PbPb data for each of the centrality bins.}
    \label{tab:systematics1}
    \begin{tabular}{lcccccc}
        \hline
        Systematic          & \pp    & \multicolumn{4}{c}{\PbPb centrality}              \\
        ~                   &        & 50--90\%  & 30--50\%  & 10--30\% & 0--10\%
        \\\hline
        Electron rejection  & 0.001 & 0.003 & 0.002 & 0.003 & 0.002
        \\
        Purity              & 0.005 & 0.008 & 0.012 & 0.026 & 0.044
        \\
        Isolation definition & 0.012 & 0.025 & 0.014 & 0.020 & 0.044
        \\
        Photon energy scale & 0.007 & 0.012 & 0.009 & 0.012 & 0.012
        \\
        Mixed-event subtraction & \NA & 0.005 & 0.009 & 0.026 & 0.095
        \\
        JEC                 & 0.063 & 0.071 & 0.098 & 0.101 & 0.100
        \\
        JER                 & 0.006 & 0.011 & 0.017 & 0.031 & 0.031
        \\
        Centrality          & \NA   & 0.014 & 0.016 & 0.012 & 0.008
        \\
        Prior choice        & 0.022 & 0.064 & 0.037 & 0.039 & 0.075
        \\
        Response matrix statistics   & 0.069 & 0.029 & 0.045 & 0.055 & 0.082
        \\
        Iteration choice    & 0.011 & 0.031 & 0.011 & 0.019 & 0.016
        \\[\cmsTabSkip]
        Total               & 0.104 & 0.125 & 0.128 & 0.142 & 0.206
        \\\hline
    \end{tabular}
\end{table*}

\begin{table*}[hbt!]
    \centering
    \topcaption{Average of the bin-by-bin \ddr absolute, symmetric systematic uncertainties for jets with $60 < \ptj < 100\GeV$ in \pp data as well as in \PbPb data for each of the centrality bins.}
    \label{tab:systematics2}
    \begin{tabular}{lcccccc}
        \hline
        Systematic          & \pp    & \multicolumn{4}{c}{\PbPb centrality}              \\
        ~                   &        & 50--90\%  & 30--50\%  & 10--30\% & 0--10\%
        \\\hline
        Electron rejection  & 0.000 & 0.001 & 0.002 & 0.001 & 0.000
        \\
        Purity              & 0.010 & 0.018 & 0.020 & 0.012 & 0.012
        \\
        Isolation definition & 0.014 & 0.017 & 0.013 & 0.012 & 0.008
        \\
        Photon energy scale & 0.009 & 0.016 & 0.009 & 0.014 & 0.012
        \\
        Mixed-event subtraction & \NA   & 0.001 & 0.002 & 0.004 & 0.001
        \\
        JEC                 & 0.058 & 0.081 & 0.087 & 0.068 & 0.070
        \\
        JER                 & 0.005 & 0.002 & 0.006 & 0.012 & 0.015
        \\
        Centrality          & \NA   & 0.012 & 0.009 & 0.005 & 0.003
        \\
        Prior choice        & 0.016 & 0.049 & 0.020 & 0.016 & 0.041
        \\
        Response matrix statistics   & 0.040 & 0.018 & 0.021 & 0.019 & 0.021
        \\
        Iteration choice    & 0.007 & 0.035 & 0.008 & 0.006 & 0.008
        \\[\cmsTabSkip]
        Total               & 0.078 & 0.113 & 0.100 & 0.081 & 0.090
        \\\hline
    \end{tabular}
\end{table*}

\emph{Response matrix statistics:} The response matrices used for unfolding are constructed with simulations that have limited event count. As a result, there are statistical uncertainties associated with each response matrix bin. To account for the uncertainty in the final result due to the limited number of simulated events, we unfold with 50 alternative response matrices, for which each bin is varied randomly according to its statistical uncertainty. The standard deviation for each bin of the unfolded \ddr spectra is taken as the systematic uncertainty. This systematic uncertainty is moderately large in \PbPb collisions and dominant in \pp collisions due to the smaller size of the available simulation samples.

\emph{Iteration choice:} The unfolding is performed with the D'Agostini iterative method with early stopping, where the stopping iteration is chosen to minimize the MSE. Theory inputs, unrelated to the \PYTHIA simulation used for the response matrices, are forward folded with the response matrix and used to generate 1000 pseudo-experiments, which then are unfolded at each iteration parameter and used to calculate the bias and the variance. However, using different theory inputs to generate the pseudo-experiments can alter the ``optimal'' iteration choice. To account for this uncertainty, the maximum difference in the \ddr spectra from the most extreme possible regularization strengths is taken as the systematic uncertainty.

The total systematic uncertainty is calculated by summing the symmetrized contributions from each source listed above in quadrature. The maximum value of the deviation is taken if the sources are mutually exclusive up/down variations of the same quantity, such as for the photon purity, JEC, centrality, and iteration choice systematics. When the ratio of \PbPb to \pp is constructed, the systematic uncertainties are taken as completely uncorrelated and added in quadrature, since the data were taken in different years. The dominant systematics in both \pp and \PbPb collisions are the JEC, the unfolding prior, and the response matrix statistics. Additionally, the mixed-event background subtraction systematic is significant in central \PbPb collisions for jet $\pt < 60\GeV$.

\section{Results}

The angular separation between jet axes determined with the \Escheme and the WTA scheme, \ddr, offers insights into the effects of momentum broadening caused by the medium in \PbPb collisions. Studying photon-tagged jet events yields a quark-enriched sample; as a result, confounding effects from differences in quark and gluon jet quenching are smaller than they are in dijet measurements. Furthermore, the jet survivor bias is reduced in photon-tagged jet measurements compared to inclusive jet measurements. Photons are unmodified by the QGP medium, and so the photon ``tags'' the initial transverse momentum from the hard scattering. By selecting on the photon \pt, we can compare hard scatterings in \pp and \PbPb collisions with approximately the same initial energy.

However, some jet survivor bias still remains, as jets lose energy due to interactions with the QGP and can subsequently fall below the minimum jet \pt threshold. This remaining jet survivor bias effect is studied by dividing the measurement into two jet \pt intervals: $30 < \ptj < 60\GeV$ and $60 < \ptj < 100\GeV$. The 60--100\GeV sample contains jets that are high energy and not quenched enough to fall below a 60\GeV threshold, and the 30--60\GeV sample captures jets which are more quenched or otherwise have energy below the minimum photon energy. The two jet \pt intervals do not show a strong correlation with the tagging photon \pt, and the lower jet \pt interval has, on average, a smaller $\xjg = \ptj / \ptg$~\cite{Sirunyan:2017qhf}.

\begin{figure*}[h!t]
    \centering
    \includegraphics[width=1.0\textwidth]{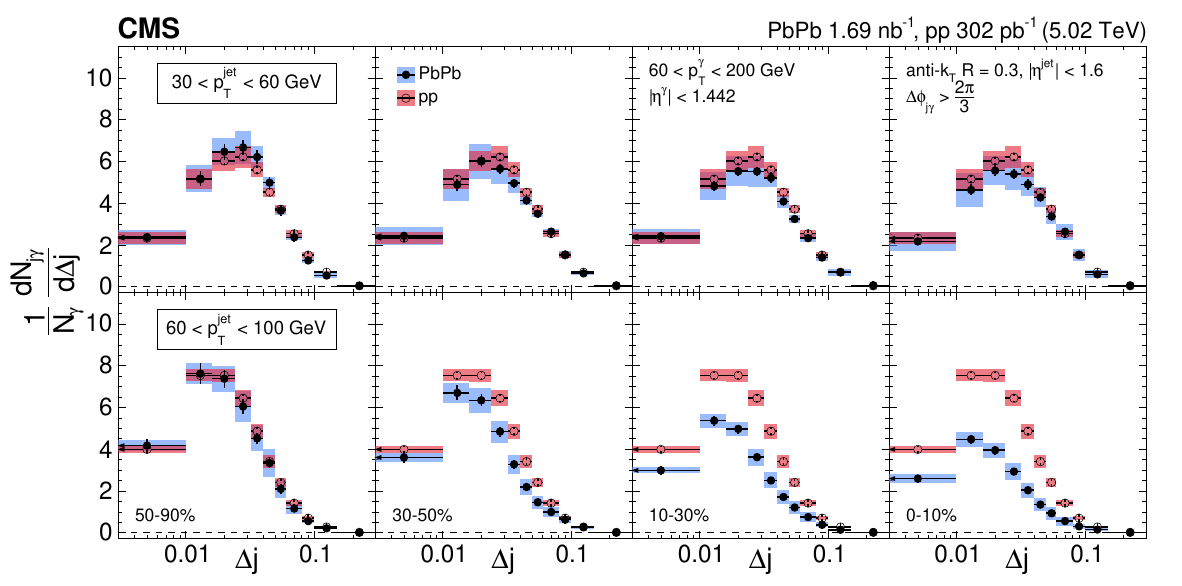}
    \caption{Photon-tagged jet axis decorrelation \ddr in \pp collisions (red boxes, open circles) and \PbPb collisions (blue boxes, filled circles), normalized per photon. The columns show different \PbPb centralities, while the top (bottom) row shows low (high) jet \pt. The leftmost bins extend down to $\ddr = 0$. The shaded boxes represent systematic uncertainties, the vertical bars indicate statistical uncertainties, and the horizontal bars indicate the bin width.}
    \label{fig:ddr_pp_aa}
\end{figure*}

Figure~\ref{fig:ddr_pp_aa} presents the photon-tagged \ddr spectra, normalized by the number of photons and corrected for detector effects, for \pp and \PbPb collisions with $30 < \ptj < 60\GeV$ and $60 < \ptj < 100\GeV$. The \PbPb collision data are presented in four centrality intervals: 0--10\%, 10--30\%, 30--50\%, and 50--90\%, and compared with the same \pp collision data. The leftmost bins extend to $\ddr = 0$, which is cut off in the plots due to the logarithmic scale on the horizontal axis. A distinct dip is observed towards $\ddr = 0$, which arises from the rarity of the $E$-- and WTA schemes yielding the same direction when multiple particles are present in the jet. The spectra reach a maximum between $0.01 < \ddr < 0.04$, followed by a steep decline as a function of \ddr. Higher-\pt jets, from 60--100\GeV, are shown to have a narrower \ddr peak with its maximum value at smaller \ddr in both \pp and \PbPb collisions compared to lower-\pt jets, from 30--60\GeV.

Figure~\ref{fig:ddr_pp_aa} also shows that the yield of $\ptj > 60\GeV$ jets is depleted as \PbPb collisions become more central, due to jet quenching. There is not as much depletion of jets below 60\GeV, since the jet \pt spectrum below 60\GeV is relatively flat due to the $60 < \ptg < 200\GeV$ photon tag requirement. Therefore, any jets that lose energy and fall below the 30\GeV \ptj threshold are likely balanced out by an approximately equal number of higher-\pt jets that have lost energy and end up in the 30--60\GeV interval.

\begin{figure*}[hbtp]
    \centering
    \includegraphics[width=1.0\textwidth]{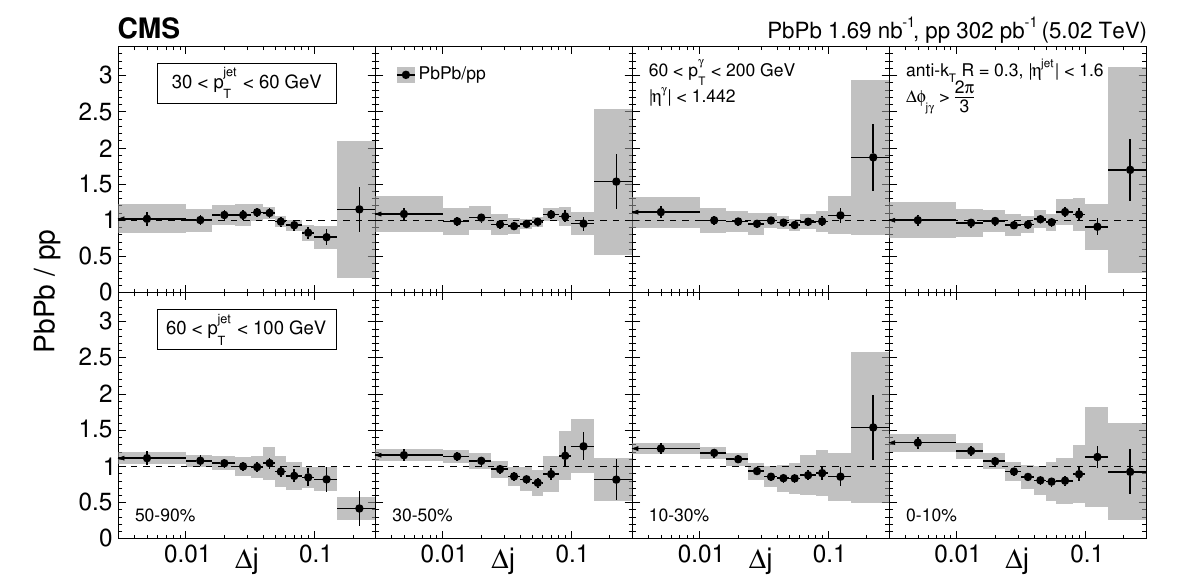}
    \caption{Ratios of photon-tagged jet \ddr spectra in \PbPb and \pp collisions normalized per photon-jet pair. The columns show different \PbPb centralities, while the top (bottom) row shows low (high) jet \pt. The leftmost bins extend down to $\ddr = 0$. The shaded boxes represent total systematic uncertainties, the vertical bars indicate statistical uncertainties, and the horizontal bars indicate the bin width.}
    \label{fig:ddr_ratio}
\end{figure*}

To quantitatively assess potential differences between the \pp and \PbPb datasets, Fig.~\ref{fig:ddr_ratio} displays the ratios of the \PbPb spectra to the \pp spectrum for each centrality interval, each now normalized by the number of photon+jet pairs. For lower \pt jets, which include more quenched jets, the ratios are found to be consistent with unity within the quoted statistical and systematic uncertainties. For higher-\pt jets, which are generally less quenched, the ratios show a hint of narrowing at small \ddr that becomes increasingly significant in central \PbPb collisions.

The survivor bias from the 60\GeV minimum jet \pt threshold may cause this narrowing. Wider jets tend to be more strongly quenched by the medium, due to stronger interaction strength arising from the resolution of more color charges~\cite{Casalderrey-Solana:2012evi}. This preferential quenching of wide jets, in combination with a steeply falling jet \pt spectrum above 60\GeV, may cause a greater proportion of narrow jets to remain in the 60--100\GeV jet \pt interval. Although wide, higher-\pt jets will also lose energy and migrate into this interval, there will be much fewer due to the increasing rarity of jets at high \pt. Additionally, all jets tend to be narrower on average as jet \pt increases, due to the larger boost, so feed-down from higher-\pt scatterings could also contribute to the narrowing more directly.

The results can be compared with the inclusive charged jet measurements reported by the ALICE Collaboration~\cite{ALICE:2023dwg}. These measurements studied inclusive $R = 0.2$ charged particle jets in two jet \pt intervals: $40 < \ptj < 60\GeV$ and $60 < \ptj < 80\GeV$. In those measurements, the normalized \PbPb to \pp spectra ratios show significant narrowing in both jet \pt intervals, reaching values as high as 1.5 at small \ddr. The CMS measurement also sees signs of narrowing at small \ddr in the $60 < \ptj < 100\GeV$ interval, similar to what was observed in the ALICE measurement. In both measurements, the narrowing is likely due to jet survivor bias effects.

However, the CMS measurement does not see any signs of narrowing in the lower \ptj interval. This lower \ptj interval, from 30--60\GeV, still has remaining bias due to the 30\GeV minimum jet \pt threshold. However, in this region the jet \pt spectrum is not as steeply falling because of the $60 < \ptg < 200\GeV$ photon tag. Furthermore, jets in this interval are more quenched on average and may show larger relative effects from interactions with the QGP. In the CMS photon-tagged jet measurement, the ratios for lower-\pt jets are consistent with unity, which could indicate a balance of survivor bias and medium-induced modification.

To gain deeper insight into the source of these trends, the experimental data are compared to several different jet-quenching models. The ``Jet Evolution With Energy Loss'' model (\jewel 2.2.0) is a dynamic framework for jet evolution in a dense background based on perturbative QCD~\cite{Kunnawalkam_Elayavalli_2017}. The ``Hybrid model'' (\hybrid) combines a perturbative description of weakly coupled jet production and evolution with a gauge/gravity duality description of strongly coupled dynamics and soft exchanges between the jet and the medium~\cite{Casalderrey-Solana:2014bpa, Casalderrey_Solana_2017, hulcher2022sensitivity}. The ``\PYTHIA Quenched'' model (\pyquen 1.5.3) calculates accumulated energy loss via gluon radiation for each parton scattering in the expanding QGP~\cite{Lokhtin_2011}. Predictions from the \jewel and \hybrid models have previously shown good agreement with a wide range of jet observables~\cite{Casalderrey_Solana_2017, Kunnawalkam_Elayavalli_2017}. In addition, study of bin feed-in and feed-out effects induced by quenching in \jewel with kinematics matching this analysis shows that jets in the $30 < \ptj < 60 \GeV$ bin are more quenched, on average, than jets in the $60 < \ptj < 100 \GeV$ bin, showing the expected survivor bias described previously.

\begin{figure*}[h!t]
    \centering
    \includegraphics[width=1.0\textwidth]{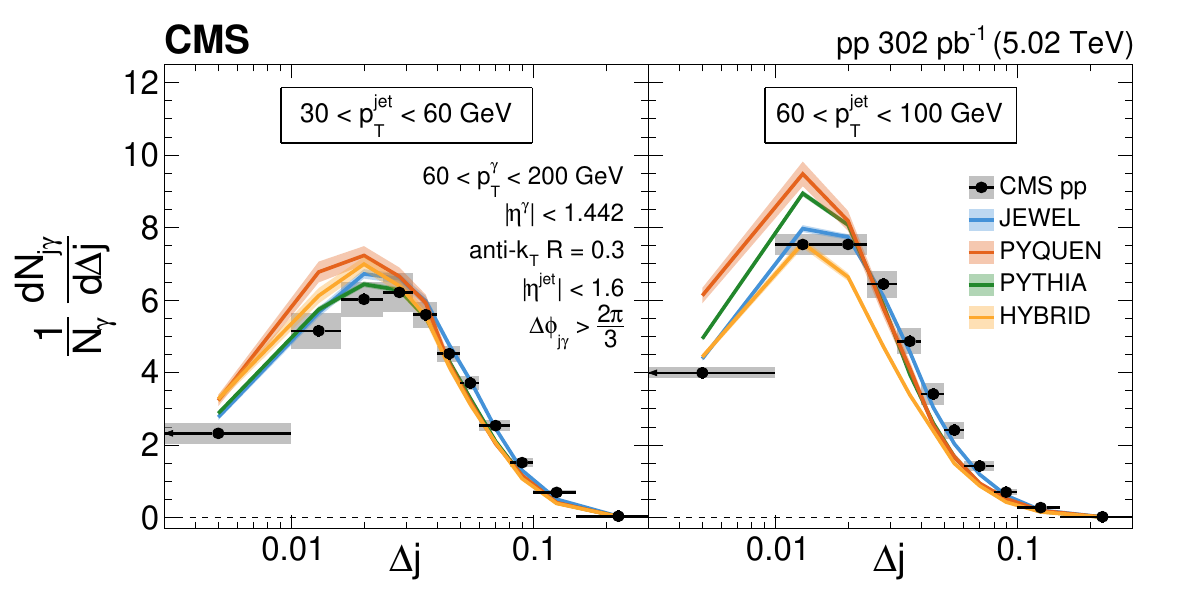}
    \caption{Photon-tagged jet axis decorrelation \ddr in \pp collisions, normalized per photon. The left (right) panel shows low (high) jet \pt. The leftmost bins extend down to $\ddr = 0$. The shaded boxes represent systematic uncertainties, statistical uncertainties are within the marker size, and the horizontal bars indicate the bin width. The results are compared to theoretical predictions from the \hybrid, \jewel, \pyquen, and \PYTHIA models. The theory bands represent the statistical uncertainties.}
    \label{fig:ddr_pp}
\end{figure*}

First, the \pp data in Fig.~\ref{fig:ddr_pp} are compared to predictions from the \hybrid, \jewel, \pyquen, and \PYTHIA models to validate the baseline model without jet quenching. For \pp collisions, the \jewel model is a modified version of \PYTHIA 6.4.25, the \pyquen model is based on \PYTHIA 6.4, and the \hybrid model is based on \PYTHIA 8.3 with the Monash tune~\cite{Skands:2014pea}, but with initial-state radiation turned on, no multiparticle interactions, and a 1\GeV cutoff for final-state radiation. The \PYTHIA prediction uses \PYTHIA 8.230 with the CP5 tune.

As the \pp data are corrected for all detector effects, comparisons are made directly with each model's generator-level output. That is, no folding-in of detector resolution is necessary to make a valid comparison. The \jewel model provides a good description of the \pp data across the \ddr spectrum. However, both \pyquen and \PYTHIA predict a narrower \ddr distribution than observed in data, overpredicting at small \ddr and underpredicting at large \ddr. The \hybrid model undershoots the tail region at large \ddr, especially at high \ptj.

\begin{figure*}[h!t]
    \centering
    \includegraphics[width=1.0\textwidth]{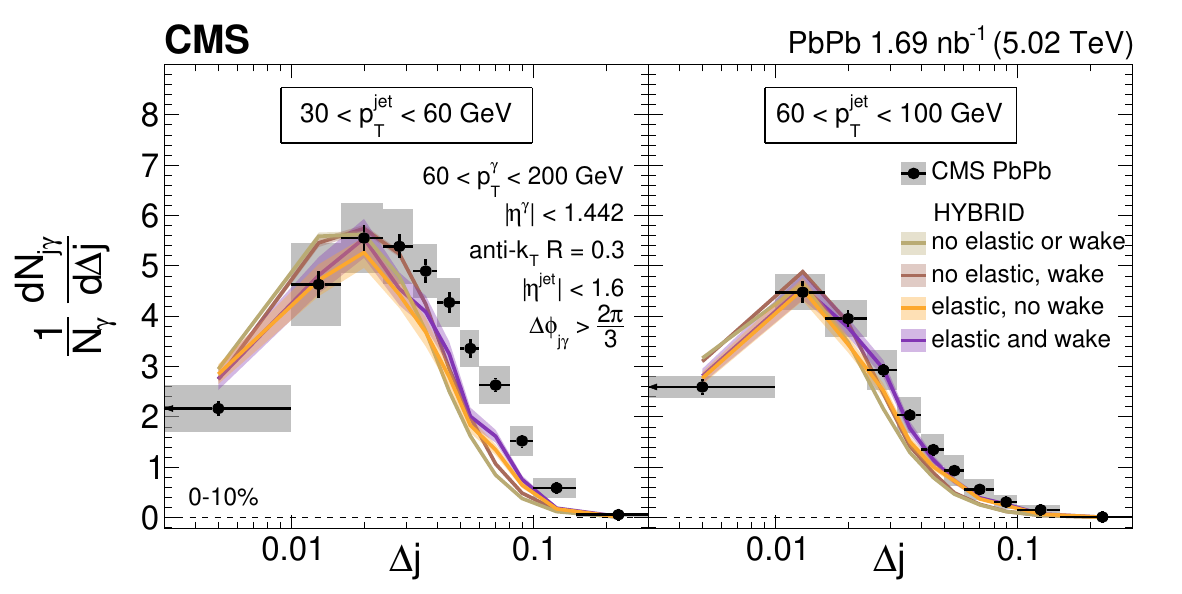}\\
    \includegraphics[width=1.0\textwidth]{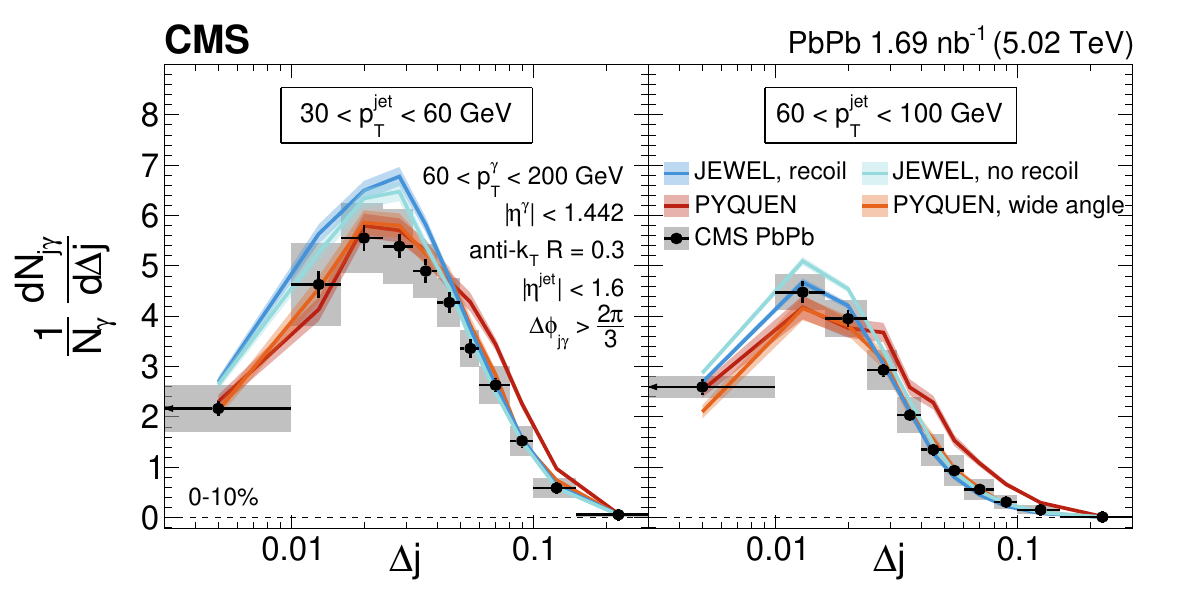}
    \caption{Photon-tagged jet axis decorrelation \ddr in \PbPb collisions with 0--10\% centrality, normalized per photon. The left (right) column shows low (high) jet \pt while the top (bottom) row shows comparisons to predictions from the \hybrid (\jewel and \pyquen) models. The leftmost bins extend down to $\ddr = 0$. The shaded boxes represent systematic uncertainties, the vertical bars indicate statistical uncertainties, and the horizontal bars indicate the bin width. The theory bands represent the statistical uncertainties.}
    \label{fig:ddr_aa}
\end{figure*}

Figure~\ref{fig:ddr_aa} shows the \ddr distribution in central \PbPb collisions compared to the \jewel, \pyquen, and \hybrid predictions, normalized per photon. The upper plots present four different curves from the \hybrid model: one showing the full model incorporating both elastic scattering and the contribution from the QGP wake, one with the wake contribution turned off, one with the elastic scattering turned off, and one with neither wake nor elastic scattering contributions. The \hybrid model undershoots the tail region of the \ddr spectra for low-\pt jets, but agrees well with data for high-\pt jets. The \hybrid predictions show the \ddr observable is sensitive to both wake and elastic scattering effects, and the inclusion of both effects yields the best description of data.

The lower plots in Fig.~\ref{fig:ddr_aa} show four predictions from \jewel and \pyquen models: the \jewel model with and without the inclusion of medium recoil effects, and the \pyquen model with and without additional wide-angle radiation. The \jewel model provides a good description of the \PbPb data at high \ptj, but overpredicts the jet yields at low \ptj. The \jewel predictions show that the inclusion of medium recoil effects, which can form a wake around the jet, alters the yield of recoiling jets. However, the shape of the distribution is largely unaffected. The \pyquen model is compatible with data only when wide-angle radiation is included in the jet-quenching model, indicating the importance of radiative energy loss.

\begin{figure*}[hbtp!]
    \centering
    \includegraphics[width=1.0\textwidth]{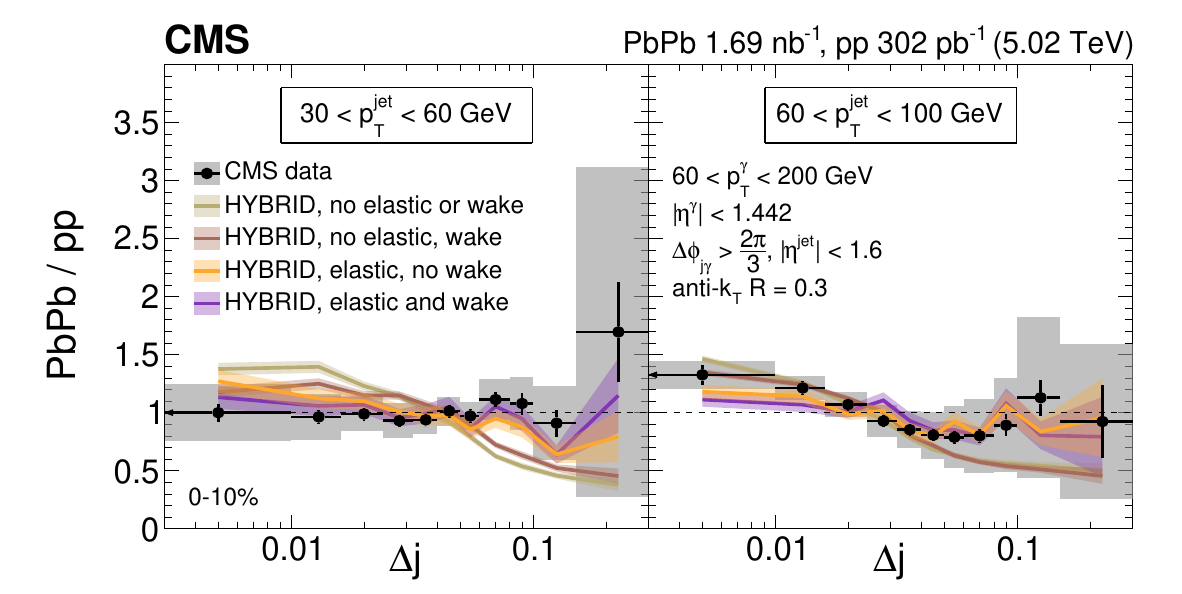}\\
    \includegraphics[width=1.0\textwidth]{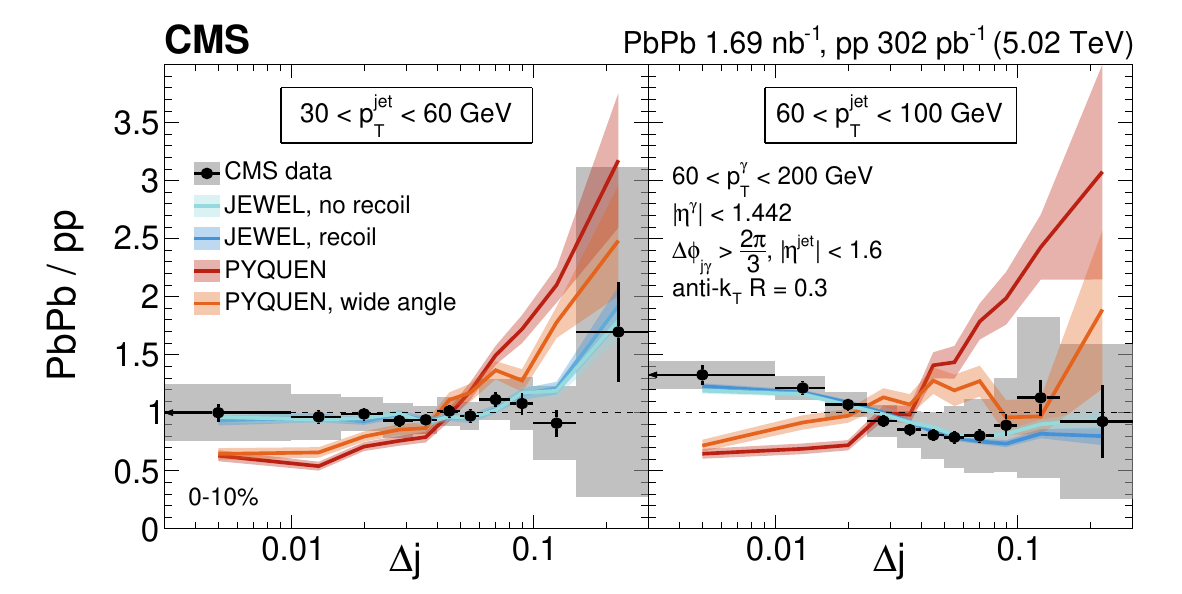}
    \caption{Ratios of photon-tagged jet axis decorrelation \ddr in \PbPb and \pp collisions normalized per photon-jet pair, shown for centrality 0--10\%. The left (right) column shows low (high) jet \pt while the top (bottom) row shows comparisons to predictions from the \hybrid (\jewel and \pyquen) models. The leftmost bins extend down to $\ddr = 0$. The shaded boxes represent systematic uncertainties, the vertical bars indicate statistical uncertainties, and the horizontal bars indicate the bin width. The theory bands represent the statistical uncertainties.}
    \label{fig:ddr_ratio_theory}
\end{figure*}

Figure~\ref{fig:ddr_ratio_theory} shows the ratios of central \PbPb to \pp data, now normalized per photon-jet pair, compared to predictions from theoretical models. The upper plots present four different predictions from the \hybrid model, which variably include elastic scattering and QGP wake effects. The curves with and without the wake contribution are now very similar, which contrasts with the differences seen in Fig.~\ref{fig:ddr_aa}. This change demonstrates that the shape of the \ddr observable is relatively insensitive to the QGP wake contribution, but the wake does slightly affect the quantity of recoiling jets in the acceptance region. In contrast, the shape of the \ddr observable is highly sensitive to the inclusion of elastic scatterings of jet shower partons with QGP constituents. The experimental data are reasonably described by the \hybrid model, providing support for its validity in explaining the observed trends. Notably, the \hybrid model without elastic scattering under-predicts the \PbPb to \pp ratios at large \ddr and over-predicts them at small \ddr, suggesting the significance of elastic scattering to describe the data.

The ratios of central \PbPb to \pp data are also compared to the predictions from the \jewel and \pyquen event generators, as illustrated in the lower plots of Fig.~\ref{fig:ddr_ratio_theory}. The results from the \jewel model with and without considering medium recoil effects are found to be very close to each other, corroborating the \hybrid model predictions and indicating that the \ddr ratios are relatively insensitive to the effects of the QGP partons recoiling from the jets. The \hybrid model with wake and \jewel with recoils are both different methods of modeling the partonic energy loss and resulting medium modifications.

The \jewel model predicts an enhancement of the \PbPb to \pp ratio at small $\ddr < 0.01$ and high \ptj, which describes the data well. Predictions from the \pyquen model are also presented, with and without wide-angle radiation. Although the full \pyquen prediction agrees well with \PbPb data in Fig.~\ref{fig:ddr_aa}, the \pyquen \pp predictions are too narrow, as seen in Fig.~\ref{fig:ddr_pp}, and so \pyquen does not properly describe jet modification in the QGP medium. The inclusion of wide-angle radiation results in a slightly better agreement with data, but both predictions severely overestimate the medium-induced broadening effects.

\section{Summary}

Measurements of the photon-tagged jet axis decorrelation (\ddr), between jet axes defined by the energy-weight (\Escheme) and the winner-take-all (WTA) schemes, are reported for the first time. These measurements are performed at a nucleon-nucleon center-of-mass energy of $\sqrtsNN = 5.02\TeV$, using proton-proton (\pp) and lead-lead (\PbPb) data sets recorded by the CMS experiment at the CERN LHC in 2017 and 2018, respectively.

This study uses comparisons between \PbPb and \pp collisions to investigate the in-medium modification of the \ddr distribution. The corrected \ddr spectra are analyzed in four \PbPb centrality intervals: 0--10\%, 10--30\%, 30--50\%, and 50--90\%. The use of photon-tagged jet events reduces the jet survivor bias and allows a focus on a quark-enriched sample with the initial transverse momentum tagged by the photon with energy between 60 and 200\GeV. This tagging requirement reduces competing effects compared to inclusive jet-based measurements, allowing for the investigation of jet survivor bias effects.

The data are divided into lower- and higher-\pt jet intervals to explore the effects of jet survivor bias. In contrast to the narrowing effects observed in the ALICE inclusive charged jet measurement in \PbPb collisions, the CMS results show that the \PbPb and \pp spectra for $30 < \ptj < 60\GeV$ are consistent. However, the higher-\pt jet sample of $60 < \ptj < 100\GeV$, which is more affected by jet survivor bias, displays similar signs of narrowing at small \ddr as reported by the ALICE experiment.

The \jewel model offers a reasonable description of the experimental data and suggests that the \ddr observable is relatively insensitive to the quark-gluon plasma (QGP) medium recoil effects. The broadening predicted by \jewel at large \ddr and low \ptj, as well as the narrowing at small \ddr and high \ptj in the most central collisions, both align with the data. Similarly, the \hybrid model indicates that \ddr is relatively unaffected by the QGP wake contribution, but is highly sensitive to elastic scattering of partons with the QGP. In contrast, the \pyquen model overestimates medium-induced broadening effects, with only slight improvement when wide-angle radiation is included. These new measurements of the photon-tagged jet axis decorrelation provide deeper insights into the mechanisms of jet quenching in the QGP.

\begin{acknowledgments}
We congratulate our colleagues in the CERN accelerator departments for the excellent performance of the LHC and thank the technical and administrative staffs at CERN and at other CMS institutes for their contributions to the success of the CMS effort. In addition, we gratefully acknowledge the computing centers and personnel of the Worldwide LHC Computing Grid and other centers for delivering so effectively the computing infrastructure essential to our analyses. Finally, we acknowledge the enduring support for the construction and operation of the LHC, the CMS detector, and the supporting computing infrastructure provided by the following funding agencies: SC (Armenia), BMBWF and FWF (Austria); FNRS and FWO (Belgium); CNPq, CAPES, FAPERJ, FAPERGS, and FAPESP (Brazil); MES and BNSF (Bulgaria); CERN; CAS, MoST, and NSFC (China); MINCIENCIAS (Colombia); MSES and CSF (Croatia); RIF (Cyprus); SENESCYT (Ecuador); ERC PRG and PSG, TARISTU24-TK10 and MoER TK202 (Estonia); Academy of Finland, MEC, and HIP (Finland); CEA and CNRS/IN2P3 (France); SRNSF (Georgia); BMFTR, DFG, and HGF (Germany); GSRI (Greece); NKFIH (Hungary); DAE and DST (India); IPM (Iran); SFI (Ireland); INFN (Italy); MSIT and NRF (Republic of Korea); MES (Latvia); LMTLT (Lithuania); MOE and UM (Malaysia); BUAP, CINVESTAV, CONACYT, LNS, SEP, and UASLP-FAI (Mexico); MOS (Montenegro); MBIE (New Zealand); PAEC (Pakistan); MES, NSC, and NAWA (Poland); FCT (Portugal);  MESTD (Serbia); MICIU/AEI and PCTI (Spain); MOSTR (Sri Lanka); Swiss Funding Agencies (Switzerland); MST (Taipei); MHESI (Thailand); TUBITAK and TENMAK (T\"{u}rkiye); NASU (Ukraine); STFC (United Kingdom); DOE and NSF (USA).

\hyphenation{Rachada-pisek} Individuals have received support from the Marie-Curie program and the European Research Council and Horizon 2020 Grant, contract Nos.\ 675440, 724704, 752730, 758316, 765710, 824093, 101115353, 101002207, 101001205, and COST Action CA16108 (European Union); the Leventis Foundation; the Alfred P.\ Sloan Foundation; the Alexander von Humboldt Foundation; the Science Committee, project no. 22rl-037 (Armenia); the Fonds pour la Formation \`a la Recherche dans l'Industrie et dans l'Agriculture (FRIA) and Fonds voor Wetenschappelijk Onderzoek contract No. 1228724N (Belgium); the Beijing Municipal Science \& Technology Commission, No. Z191100007219010, the Fundamental Research Funds for the Central Universities, the Ministry of Science and Technology of China under Grant No. 2023YFA1605804, the Natural Science Foundation of China under Grant No. 12535004, and USTC Research Funds of the Double First-Class Initiative No.\ YD2030002017 (China); the Ministry of Education, Youth and Sports (MEYS) of the Czech Republic; the Shota Rustaveli National Science Foundation, grant FR-22-985 (Georgia); the Deutsche Forschungsgemeinschaft (DFG), among others, under Germany's Excellence Strategy -- EXC 2121 ``Quantum Universe" -- 390833306, and under project number 400140256 - GRK2497; the Hellenic Foundation for Research and Innovation (HFRI), Project Number 2288 (Greece); the Hungarian Academy of Sciences, the New National Excellence Program - \'UNKP, the NKFIH research grants K 131991, K 133046, K 138136, K 143460, K 143477, K 146913, K 146914, K 147048, 2020-2.2.1-ED-2021-00181, TKP2021-NKTA-64, and 2025-1.1.5-NEMZ\_KI-2025-00004 (Hungary); the Council of Science and Industrial Research, India; ICSC -- National Research Center for High Performance Computing, Big Data and Quantum Computing, FAIR -- Future Artificial Intelligence Research, and CUP I53D23001070006 (Mission 4 Component 1), funded by the NextGenerationEU program (Italy); the Latvian Council of Science; the Ministry of Education and Science, project no. 2022/WK/14, and the National Science Center, contracts Opus 2021/41/B/ST2/01369, 2021/43/B/ST2/01552, 2023/49/B/ST2/03273, and the NAWA contract BPN/PPO/2021/1/00011 (Poland); the Funda\c{c}\~ao para a Ci\^encia e a Tecnologia (Portugal); the National Priorities Research Program by Qatar National Research Fund; MICIU/AEI/10.13039/501100011033, ERDF/EU, ``European Union NextGenerationEU/PRTR", and Programa Severo Ochoa del Principado de Asturias (Spain); the Chulalongkorn Academic into Its 2nd Century Project Advancement Project, the National Science, Research and Innovation Fund program IND\_FF\_68\_369\_2300\_097, and the Program Management Unit for Human Resources \& Institutional Development, Research and Innovation, grant B39G680009 (Thailand); the Eric \& Wendy Schmidt Fund for Strategic Innovation through the CERN Next Generation Triggers project under grant agreement number SIF-2023-004; the Kavli Foundation; the Nvidia Corporation; the SuperMicro Corporation; the Welch Foundation, contract C-1845; and the Weston Havens Foundation (USA).
\end{acknowledgments}\section*{Data availability} Release and preservation of data used by the CMS Collaboration as the basis for publications is guided by the  \href{https://doi.org/10.7483/OPENDATA.CMS.1BNU.8V1W}{CMS data preservation, re-use and open access policy}.

\bibliography{auto_generated}

\providecommand{\href}[2]{#2}\begingroup\raggedright\begin{thebibliography}{10}%
\makeatletter
\providecommand{\hrefCMSnoop }[0]{\@secondoftwo}%
\makeatother
\providecommand{\doi}{\texttt{doi:}\begingroup \urlstyle{tt}\Url}

\bibitem{Collins:1974ky}
\hrefCMSnoop {}{J.~C. Collins and M.~J. Perry, ``Superdense matter: Neutrons or
  asymptotically free quarks?'',} \textit{ Phys. Rev. Lett.} \textbf{ 34}
  (1975) 1353,
\href{http://dx.doi.org/10.1103/PhysRevLett.34.1353}{\doi{10.1103/PhysRevLett.34.1353}}.

\bibitem{Karsch:1995sy}
\hrefCMSnoop {}{F.~Karsch, ``{The phase transition to the quark gluon plasma:
  Recent results from lattice calculations}'',} \textit{ Nucl. Phys. A}
  \textbf{ 590} (1995) 367,
  \href{http://dx.doi.org/10.1016/0375-9474(95)00248-Y}{\doi{10.1016/0375-9474(95)00248-Y}},
\href{http://www.arXiv.org/abs/hep-lat/9503010}{\texttt{arXiv:hep-lat/9503010}}.

\bibitem{Busza:2018rrf}
\hrefCMSnoop {}{W.~Busza, K.~Rajagopal, and W.~van~der Schee, ``Heavy ion
  collisions: The big picture, and the big questions'',} \textit{ Ann. Rev.
  Nucl. Part. Sci.} \textbf{ 68} (2018) 339,
  \href{http://dx.doi.org/10.1146/annurev-nucl-101917-020852}{\doi{10.1146/annurev-nucl-101917-020852}},
\href{http://www.arXiv.org/abs/1802.04801}{\texttt{arXiv:1802.04801}}.

\bibitem{Bjorken:1982qr}
\hrefCMSnoop {}{J.~D. Bjorken, ``{Highly relativistic nucleus-nucleus
  collisions: the central rapidity region}'',} \textit{ Phys. Rev. D} \textbf{
  27} (1983) 140,
\href{http://dx.doi.org/10.1103/PhysRevD.27.140}{\doi{10.1103/PhysRevD.27.140}}.

\bibitem{Appel:1985dq}
\hrefCMSnoop {}{D.~A. Appel, ``{Jets as a probe of quark-gluon plasmas}'',}
  \textit{ Phys. Rev. D} \textbf{ 33} (1986) 717,
\href{http://dx.doi.org/10.1103/PhysRevD.33.717}{\doi{10.1103/PhysRevD.33.717}}.

\bibitem{Blaizot:1986ma}
\hrefCMSnoop {}{J.~P. Blaizot and L.~D. McLerran, ``{Jets in expanding
  quark-gluon plasmas}'',} \textit{ Phys. Rev. D} \textbf{ 34} (1986) 2739,
\href{http://dx.doi.org/10.1103/PhysRevD.34.2739}{\doi{10.1103/PhysRevD.34.2739}}.

\bibitem{Gyulassy:1990ye}
\hrefCMSnoop {}{M.~Gyulassy and M.~Pl{\"u}mer, ``{Jet quenching in dense
  matter}'',} \textit{ Phys. Lett. B} \textbf{ 243} (1990) 432,
\href{http://dx.doi.org/10.1016/0370-2693(90)91409-5}{\doi{10.1016/0370-2693(90)91409-5}}.

\bibitem{Wang:1991xy}
\hrefCMSnoop {}{X.-N. Wang and M.~Gyulassy, ``{Gluon shadowing and jet
  quenching in AA collisions at $\sqrt{s} = 200A$ GeV}'',} \textit{ Phys. Rev.
  Lett.} \textbf{ 68} (1992) 1480,
\href{http://dx.doi.org/10.1103/PhysRevLett.68.1480}{\doi{10.1103/PhysRevLett.68.1480}}.

\bibitem{Baier:1996sk}
R.~Baier\hrefCMSnoop {}{ { et~al.}, ``{Radiative energy loss and \pt-broadening
  of high energy partons in nuclei}'',} \textit{ Nucl. Phys. B} \textbf{ 484}
  (1997) 265,
  \href{http://dx.doi.org/10.1016/S0550-3213(96)00581-0}{\doi{10.1016/S0550-3213(96)00581-0}},
\href{http://www.arXiv.org/abs/hep-ph/9608322}{\texttt{arXiv:hep-ph/9608322}}.

\bibitem{Zakharov:1997uu}
\hrefCMSnoop {}{B.~G. Zakharov, ``{Radiative energy loss of high-energy quarks
  in finite-size nuclear matter and quark-gluon plasma}'',} \textit{ JETP
  Lett.} \textbf{ 65} (1997) 615,
  \href{http://dx.doi.org/10.1134/1.567389}{\doi{10.1134/1.567389}},
\href{http://www.arXiv.org/abs/hep-ph/9704255}{\texttt{arXiv:hep-ph/9704255}}.

\bibitem{Connors:2017ptx}
\hrefCMSnoop {}{M.~Connors, C.~Nattrass, R.~Reed, and S.~Salur, ``{Jet
  measurements in heavy ion physics}'',} \textit{ Rev. Mod. Phys.} \textbf{ 90}
  (2018) 025005,
  \href{http://dx.doi.org/10.1103/RevModPhys.90.025005}{\doi{10.1103/RevModPhys.90.025005}},
  \href{http://www.arXiv.org/abs/1705.01974}{\texttt{arXiv:1705.01974}}.

\bibitem{Cunqueiro:2021wls}
\hrefCMSnoop {}{L.~Cunqueiro and A.~M. Sickles, ``{Studying the QGP with jets
  at the LHC and RHIC}'',} \textit{ Prog. Part. Nucl. Phys.} \textbf{ 124}
  (2022) 103940,
  \href{http://dx.doi.org/10.1016/j.ppnp.2022.103940}{\doi{10.1016/j.ppnp.2022.103940}},
  \href{http://www.arXiv.org/abs/2110.14490}{\texttt{arXiv:2110.14490}}.

\bibitem{Apolinario:2022vzg}
\hrefCMSnoop {}{L.~Apolinario, Y.-J. Lee, and M.~Winn, ``{Heavy quarks and jets
  as probes of the QGP}'',} \textit{ Prog. Part. Nucl. Phys.} \textbf{ 127}
  (2022) 103990,
  \href{http://dx.doi.org/10.1016/j.ppnp.2022.103990}{\doi{10.1016/j.ppnp.2022.103990}},
  \href{http://www.arXiv.org/abs/2203.16352}{\texttt{arXiv:2203.16352}}.

\bibitem{CMS:2024krd}
\hrefCMSnoop {}{{CMS Collaboration}, ``{Overview of high-density QCD studies
  with the CMS experiment at the LHC}'',} \textit{ Phys. Rept.} \textbf{ 1115}
  (2025) 219,
  \href{http://dx.doi.org/10.1016/j.physrep.2024.11.007}{\doi{10.1016/j.physrep.2024.11.007}},
  \href{http://www.arXiv.org/abs/2405.10785}{\texttt{arXiv:2405.10785}}.

\bibitem{Adams:2003kv}
\hrefCMSnoop {}{{STAR} Collaboration, ``Transverse-momentum and
  collision-energy dependence of high-\pt hadron suppression in {AuAu}
  collisions at ultrarelativistic energies'',} \textit{ Phys. Rev. Lett.}
  \textbf{ 91} (2003) 172302,
  \href{http://dx.doi.org/10.1103/PhysRevLett.91.172302}{\doi{10.1103/PhysRevLett.91.172302}},
\href{http://www.arXiv.org/abs/nucl-ex/0305015}{\texttt{arXiv:nucl-ex/0305015}}.

\bibitem{Adare:2008qa}
\hrefCMSnoop {}{{PHENIX} Collaboration, ``Suppression pattern of neutral pions
  at high transverse momentum in {AuAu} collisions at {$\sqrtsNN = 200$\GeV}
  and constraints on medium transport coefficients'',} \textit{ Phys. Rev.
  Lett.} \textbf{ 101} (2008) 232301,
  \href{http://dx.doi.org/10.1103/PhysRevLett.101.232301}{\doi{10.1103/PhysRevLett.101.232301}},
\href{http://www.arXiv.org/abs/0801.4020}{\texttt{arXiv:0801.4020}}.

\bibitem{Abelev:2012hxa}
\hrefCMSnoop {}{{ALICE Collaboration}, ``{Centrality dependence of charged
  particle production at large transverse momentum in PbPb collisions at
  $\sqrtsNN = 2.76$\TeV}'',} \textit{ Phys. Lett. B} \textbf{ 720} (2013) 52,
  \href{http://dx.doi.org/10.1016/j.physletb.2013.01.051}{\doi{10.1016/j.physletb.2013.01.051}},
\href{http://www.arXiv.org/abs/1208.2711}{\texttt{arXiv:1208.2711}}.

\bibitem{Aad:2015wga}
\hrefCMSnoop {}{{ATLAS Collaboration}, ``{Measurement of charged-particle
  spectra in PbPb collisions at $\sqrtsNN = 2.76\TeV$ with the ATLAS detector
  at the LHC}'',} \textit{ JHEP} \textbf{ 09} (2015) 050,
  \href{http://dx.doi.org/10.1007/JHEP09(2015)050}{\doi{10.1007/JHEP09(2015)050}},
\href{http://www.arXiv.org/abs/1504.04337}{\texttt{arXiv:1504.04337}}.

\bibitem{CMS:2012aa}
\hrefCMSnoop {}{{CMS Collaboration}, ``{Study of high-$p_{\rm{T}}$ charged
  particle suppression in PbPb compared to pp collisions at
  $\sqrtsNN=2.76$\TeV}'',} \textit{ Eur. Phys. J. C} \textbf{ 72} (2012) 1945,
  \href{http://dx.doi.org/10.1140/epjc/s10052-012-1945-x}{\doi{10.1140/epjc/s10052-012-1945-x}},
\href{http://www.arXiv.org/abs/1202.2554}{\texttt{arXiv:1202.2554}}.

\bibitem{Khachatryan:2016odn}
\hrefCMSnoop {}{{CMS Collaboration}, ``{Charged-particle nuclear modification
  factors in PbPb and pPb collisions at $\sqrtsNN=5.02$\TeV}'',} \textit{ JHEP}
  \textbf{ 04} (2017) 039,
  \href{http://dx.doi.org/10.1007/JHEP04(2017)039}{\doi{10.1007/JHEP04(2017)039}},
\href{http://www.arXiv.org/abs/1611.01664}{\texttt{arXiv:1611.01664}}.

\bibitem{Chatrchyan:2011sx}
\hrefCMSnoop {}{{CMS Collaboration}, ``{Observation and studies of jet
  quenching in PbPb collisions at $\sqrtsNN = 2.76$\TeV}'',} \textit{ Phys.
  Rev. C} \textbf{ 84} (2011) 024906,
  \href{http://dx.doi.org/10.1103/PhysRevC.84.024906}{\doi{10.1103/PhysRevC.84.024906}},
  \href{http://www.arXiv.org/abs/1102.1957}{\texttt{arXiv:1102.1957}}.

\bibitem{Aad:2010bu}
\hrefCMSnoop {}{{ATLAS Collaboration}, ``Observation of a centrality-dependent
  dijet asymmetry in lead-lead collisions at {$\sqrtsNN = 2.76$\TeV} with the
  {ATLAS} detector at the {LHC}'',} \textit{ Phys. Rev. Lett.} \textbf{ 105}
  (2010) 252303,
  \href{http://dx.doi.org/10.1103/PhysRevLett.105.252303}{\doi{10.1103/PhysRevLett.105.252303}},
  \href{http://www.arXiv.org/abs/1011.6182}{\texttt{arXiv:1011.6182}}.

\bibitem{ATLAS:2014cpa}
\hrefCMSnoop {}{{ATLAS Collaboration}, ``{Centrality and rapidity dependence of
  inclusive jet production in $\sqrtsNN = 5.02$\TeV proton-lead collisions with
  the ATLAS detector}'',} \textit{ Phys. Lett. B} \textbf{ 748} (2015) 392,
  \href{http://dx.doi.org/10.1016/j.physletb.2015.07.023}{\doi{10.1016/j.physletb.2015.07.023}},
\href{http://www.arXiv.org/abs/1412.4092}{\texttt{arXiv:1412.4092}}.

\bibitem{Adam:2015doa}
\hrefCMSnoop {}{{ALICE Collaboration}, ``{Measurement of jet quenching with
  semi-inclusive hadron-jet distributions in central PbPb collisions at
  $\sqrtsNN=2.76$\TeV}'',} \textit{ JHEP} \textbf{ 09} (2015) 170,
  \href{http://dx.doi.org/10.1007/JHEP09(2015)170}{\doi{10.1007/JHEP09(2015)170}},
\href{http://www.arXiv.org/abs/1506.03984}{\texttt{arXiv:1506.03984}}.

\bibitem{Khachatryan:2016jfl}
\hrefCMSnoop {}{{CMS Collaboration}, ``{Measurement of inclusive jet cross
  sections in pp and PbPb collisions at $\sqrtsNN=2.76\TeV$}'',} \textit{ Phys.
  Rev. C} \textbf{ 96} (2017) 015202,
  \href{http://dx.doi.org/10.1103/PhysRevC.96.015202}{\doi{10.1103/PhysRevC.96.015202}},
\href{http://www.arXiv.org/abs/1609.05383}{\texttt{arXiv:1609.05383}}.

\bibitem{Adam:2015ewa}
\hrefCMSnoop {}{{ALICE Collaboration}, ``{Measurement of jet suppression in
  central PbPb collisions at $\sqrtsNN = 2.76\TeV$}'',} \textit{ Phys. Lett. B}
  \textbf{ 746} (2015) 1,
  \href{http://dx.doi.org/10.1016/j.physletb.2015.04.039}{\doi{10.1016/j.physletb.2015.04.039}},
\href{http://www.arXiv.org/abs/1502.01689}{\texttt{arXiv:1502.01689}}.

\bibitem{Adamczyk:2016fqm}
\hrefCMSnoop {}{{STAR} Collaboration, ``{Dijet imbalance measurements in AuAu
  and pp collisions at $\sqrtsNN = 200\GeV$ at STAR}'',} \textit{ Phys. Rev.
  Lett.} \textbf{ 119} (2017) 062301,
  \href{http://dx.doi.org/10.1103/PhysRevLett.119.062301}{\doi{10.1103/PhysRevLett.119.062301}},
\href{http://www.arXiv.org/abs/1609.03878}{\texttt{arXiv:1609.03878}}.

\bibitem{hulcher2022sensitivity}
\hrefCMSnoop {}{Z.~Hulcher, D.~Pablos, and K.~Rajagopal, ``{Sensitivity of jet
  observables to the presence of quasi-particles in QGP}'',} in \textit{ Proc.
  29th Int. Conf. on Utrarelativistic Nucleus-Nucleus Collisions (QM 2022),
  Krakow, Poland, 4-10 April, 2022}.
\newblock 2022.
\newblock
  \href{http://www.arXiv.org/abs/2208.13593}{\texttt{arXiv:2208.13593}}.

\bibitem{Cal:2020}
\hrefCMSnoop {}{{CMS Collaboration}, ``Calculating the angle between jet
  axes'',} \textit{ JHEP} \textbf{ 04} (2020) 211,
  \href{http://dx.doi.org/10.1007/jhep04(2020)211}{\doi{10.1007/jhep04(2020)211}},
  \href{http://www.arXiv.org/abs/1911.06840}{\texttt{arXiv:1911.06840}}.

\bibitem{Cacciari:2008gp}
\hrefCMSnoop {}{M.~Cacciari, G.~P. Salam, and G.~Soyez, ``{The anti-\kt jet
  clustering algorithm}'',} \textit{ JHEP} \textbf{ 04} (2008) 063,
  \href{http://dx.doi.org/10.1088/1126-6708/2008/04/063}{\doi{10.1088/1126-6708/2008/04/063}},
\href{http://www.arXiv.org/abs/0802.1189}{\texttt{arXiv:0802.1189}}.

\bibitem{Bertolini:2013iqa}
\hrefCMSnoop {}{D.~Bertolini, T.~Chan, and J.~Thaler, ``{Jet observables
  without jet algorithms}'',} \textit{ JHEP} \textbf{ 04} (2014) 013,
  \href{http://dx.doi.org/10.1007/JHEP04(2014)013}{\doi{10.1007/JHEP04(2014)013}},
  \href{http://www.arXiv.org/abs/1310.7584}{\texttt{arXiv:1310.7584}}.

\bibitem{Brewer:2021hmh}
\hrefCMSnoop {}{J.~Brewer, Q.~Brodsky, and K.~Rajagopal, ``{Disentangling jet
  modification in jet simulations and in Z+jet data}'',} \textit{ JHEP}
  \textbf{ 02} (2022) 175,
  \href{http://dx.doi.org/10.1007/JHEP02(2022)175}{\doi{10.1007/JHEP02(2022)175}},
  \href{http://www.arXiv.org/abs/2110.13159}{\texttt{arXiv:2110.13159}}.

\bibitem{ALICE:2023dwg}
\hrefCMSnoop {}{{ALICE Collaboration}, ``{Measurement of the angle between jet
  axes in PbPb collisions at $\sqrtsNN = 5.02$ TeV}'',} 2023.
  \href{http://www.arXiv.org/abs/2303.13347}{\texttt{arXiv:2303.13347}}.
  Submitted to: Phys. Rev. Lett.

\bibitem{Kartvelishvili:1995fr}
\hrefCMSnoop {}{V.~Kartvelishvili, R.~Kvatadze, and R.~Shanidze, ``{On Z and
  Z+jet production in heavy ion collisions}'',} \textit{ Phys. Lett. B}
  \textbf{ 356} (1995) 589,
  \href{http://dx.doi.org/10.1016/0370-2693(95)00865-I}{\doi{10.1016/0370-2693(95)00865-I}},
\href{http://www.arXiv.org/abs/hep-ph/9505418}{\texttt{arXiv:hep-ph/9505418}}.

\bibitem{Wang:1996yh}
\hrefCMSnoop {}{X.-N. Wang, Z.~Huang, and I.~Sarcevic, ``Jet quenching in the
  direction opposite to a tagged photon in high-energy heavy ion collisions'',}
  \textit{ Phys. Rev. Lett.} \textbf{ 77} (1996) 231,
  \href{http://dx.doi.org/10.1103/PhysRevLett.77.231}{\doi{10.1103/PhysRevLett.77.231}},
\href{http://www.arXiv.org/abs/hep-ph/9605213}{\texttt{arXiv:hep-ph/9605213}}.

\bibitem{Wang:1996pe}
\hrefCMSnoop {}{X.-N. Wang and Z.~Huang, ``{Medium-induced parton energy loss
  in $\gamma$+jet events of high-energy heavy ion collisions}'',} \textit{
  Phys. Rev. C} \textbf{ 55} (1997) 3047,
  \href{http://dx.doi.org/10.1103/PhysRevC.55.3047}{\doi{10.1103/PhysRevC.55.3047}},
\href{http://www.arXiv.org/abs/hep-ph/9701227}{\texttt{arXiv:hep-ph/9701227}}.

\bibitem{Dai:2012am}
\hrefCMSnoop {}{W.~Dai, I.~Vitev, and B.-W. Zhang, ``{Momentum imbalance of
  isolated photon-tagged jet production at RHIC and LHC}'',} \textit{ Phys.
  Rev. Lett.} \textbf{ 110} (2013) 142001,
  \href{http://dx.doi.org/10.1103/PhysRevLett.110.142001}{\doi{10.1103/PhysRevLett.110.142001}},
\href{http://www.arXiv.org/abs/1207.5177}{\texttt{arXiv:1207.5177}}.

\bibitem{Kang:2017xnc}
\hrefCMSnoop {}{Z.-B. Kang, I.~Vitev, and H.~Xing, ``{Vector-boson-tagged jet
  production in heavy ion collisions at energies available at the CERN Large
  Hadron Collider}'',} \textit{ Phys. Rev. C} \textbf{ 96} (2017) 014912,
  \href{http://dx.doi.org/10.1103/PhysRevC.96.014912}{\doi{10.1103/PhysRevC.96.014912}},
\href{http://www.arXiv.org/abs/1702.07276}{\texttt{arXiv:1702.07276}}.

\bibitem{Aad:2015lcb}
\hrefCMSnoop {}{{ATLAS Collaboration}, ``{Centrality, rapidity and transverse
  momentum dependence of isolated prompt photon production in lead-lead
  collisions at $\sqrtsNN = 2.76\TeV$ measured with the ATLAS detector}'',}
  \textit{ Phys. Rev. C} \textbf{ 93} (2016) 034914,
  \href{http://dx.doi.org/10.1103/PhysRevC.93.034914}{\doi{10.1103/PhysRevC.93.034914}},
\href{http://www.arXiv.org/abs/1506.08552}{\texttt{arXiv:1506.08552}}.

\bibitem{Chatrchyan:2012vq}
\hrefCMSnoop {}{{CMS Collaboration}, ``{Measurement of isolated photon
  production in pp and PbPb collisions at $\sqrtsNN=2.76$\TeV}'',} \textit{
  Phys. Lett. B} \textbf{ 710} (2012) 256,
  \href{http://dx.doi.org/10.1016/j.physletb.2012.02.077}{\doi{10.1016/j.physletb.2012.02.077}},
\href{http://www.arXiv.org/abs/1201.3093}{\texttt{arXiv:1201.3093}}.

\bibitem{Chatrchyan:2012nt}
\hrefCMSnoop {}{{CMS Collaboration}, ``{Study of W boson production in PbPb and
  pp collisions at $\sqrtsNN=2.76$\TeV}'',} \textit{ Phys. Lett. B} \textbf{
  715} (2012) 66,
  \href{http://dx.doi.org/10.1016/j.physletb.2012.07.025}{\doi{10.1016/j.physletb.2012.07.025}},
\href{http://www.arXiv.org/abs/1205.6334}{\texttt{arXiv:1205.6334}}.

\bibitem{Chatrchyan:2014csa}
\hrefCMSnoop {}{{CMS Collaboration}, ``{Study of Z production in PbPb and pp
  collisions at $\sqrtsNN=2.76$\TeV in the dimuon and dielectron decay
  channels}'',} \textit{ JHEP} \textbf{ 03} (2015) 022,
  \href{http://dx.doi.org/10.1007/JHEP03(2015)022}{\doi{10.1007/JHEP03(2015)022}},
\href{http://www.arXiv.org/abs/1410.4825}{\texttt{arXiv:1410.4825}}.

\bibitem{CMS:2020oen}
\hrefCMSnoop {}{{CMS Collaboration}, ``{The production of isolated photons in
  PbPb and pp collisions at $\sqrtsNN =$ 5.02 TeV}'',} \textit{ JHEP} \textbf{
  07} (2020) 116,
  \href{http://dx.doi.org/10.1007/JHEP07(2020)116}{\doi{10.1007/JHEP07(2020)116}},
  \href{http://www.arXiv.org/abs/2003.12797}{\texttt{arXiv:2003.12797}}.

\bibitem{Neufeld:2010fj}
\hrefCMSnoop {}{R.~B. Neufeld, I.~Vitev, and B.~W. Zhang, ``Physics of
  {$Z^0/\gamma^*$}-tagged jets at energies available at the {CERN Large Hadron
  Collider}'',} \textit{ Phys. Rev. C} \textbf{ 83} (2011) 034902,
  \href{http://dx.doi.org/10.1103/PhysRevC.83.034902}{\doi{10.1103/PhysRevC.83.034902}},
\href{http://www.arXiv.org/abs/1006.2389}{\texttt{arXiv:1006.2389}}.

\bibitem{Wang:2013cia}
\hrefCMSnoop {}{X.-N. Wang and Y.~Zhu, ``{Medium modification of $\gamma$ jets
  in high-energy heavy ion collisions}'',} \textit{ Phys. Rev. Lett.} \textbf{
  111} (2013) 062301,
  \href{http://dx.doi.org/10.1103/PhysRevLett.111.062301}{\doi{10.1103/PhysRevLett.111.062301}},
\href{http://www.arXiv.org/abs/1302.5874}{\texttt{arXiv:1302.5874}}.

\bibitem{Casalderrey-Solana:2014bpa}
J.~Casalderrey-Solana\hrefCMSnoop {}{ { et~al.}, ``{A hybrid strong/weak
  coupling approach to jet quenching}'',} \textit{ JHEP} \textbf{ 10} (2014)
  019,
  \href{http://dx.doi.org/10.1007/JHEP10(2014)019}{\doi{10.1007/JHEP10(2014)019}},
\href{http://www.arXiv.org/abs/1405.3864}{\texttt{arXiv:1405.3864}}.

\bibitem{Casalderrey-Solana:2015vaa}
J.~Casalderrey-Solana\hrefCMSnoop {}{ { et~al.}, ``{Predictions for boson-jet
  observables and fragmentation function ratios from a hybrid strong/weak
  coupling model for jet quenching}'',} \textit{ JHEP} \textbf{ 03} (2016) 053,
  \href{http://dx.doi.org/10.1007/JHEP03(2016)053}{\doi{10.1007/JHEP03(2016)053}},
\href{http://www.arXiv.org/abs/1508.00815}{\texttt{arXiv:1508.00815}}.

\bibitem{KunnawalkamElayavalli:2016ttl}
\hrefCMSnoop {}{R.~K. Elayavalli and K.~C. Zapp, ``{Simulating V+jet processes
  in heavy ion collisions with JEWEL}'',} \textit{ Eur. Phys. J. C} \textbf{
  76} (2016) 695,
  \href{http://dx.doi.org/10.1140/epjc/s10052-016-4534-6}{\doi{10.1140/epjc/s10052-016-4534-6}},
\href{http://www.arXiv.org/abs/1608.03099}{\texttt{arXiv:1608.03099}}.

\bibitem{CMS:2012ytf}
\hrefCMSnoop {}{{CMS Collaboration}, ``{Studies of jet quenching using isolated
  photon+jet correlations in PbPb and $pp$ collisions at $\sqrtsNN=2.76$
  TeV}'',} \textit{ Phys. Lett. B} \textbf{ 718} (2013) 773,
  \href{http://dx.doi.org/10.1016/j.physletb.2012.11.003}{\doi{10.1016/j.physletb.2012.11.003}},
  \href{http://www.arXiv.org/abs/1205.0206}{\texttt{arXiv:1205.0206}}.

\bibitem{CMS:2021otx}
\hrefCMSnoop {}{{CMS Collaboration}, ``{Using Z boson events to study
  parton-medium interactions in PbPb collisions}'',} \textit{ Phys. Rev. Lett.}
  \textbf{ 128} (2022) 122301,
  \href{http://dx.doi.org/10.1103/PhysRevLett.128.122301}{\doi{10.1103/PhysRevLett.128.122301}},
  \href{http://www.arXiv.org/abs/2103.04377}{\texttt{arXiv:2103.04377}}.

\bibitem{Sirunyan:2017qhf}
\hrefCMSnoop {}{{CMS Collaboration}, ``{Study of jet quenching with isolated
  photon+jet correlations in PbPb and pp collisions at $\sqrtsNN =$ 5.02
  TeV}'',} \textit{ Phys. Lett. B} \textbf{ 785} (2018) 14,
  \href{http://dx.doi.org/10.1016/j.physletb.2018.07.061}{\doi{10.1016/j.physletb.2018.07.061}},
\href{http://www.arXiv.org/abs/1711.09738}{\texttt{arXiv:1711.09738}}.

\bibitem{Sirunyan:2018ncy}
\hrefCMSnoop {}{{CMS Collaboration}, ``{Jet shapes of isolated photon-tagged
  jets in PbPb and pp collisions at $\sqrtsNN =$ 5.02 TeV}'',} \textit{ Phys.
  Rev. Lett.} \textbf{ 122} (2019) 152001,
  \href{http://dx.doi.org/10.1103/PhysRevLett.122.152001}{\doi{10.1103/PhysRevLett.122.152001}},
\href{http://www.arXiv.org/abs/1809.08602}{\texttt{arXiv:1809.08602}}.

\bibitem{Sirunyan:2018qec}
\hrefCMSnoop {}{{CMS Collaboration}, ``{Observation of medium-induced
  modifications of jet fragmentation in PbPb collisions at $\sqrtsNN=$ 5.02 TeV
  using isolated photon-tagged jets}'',} \textit{ Phys. Rev. Lett.} \textbf{
  121} (2018) 242301,
  \href{http://dx.doi.org/10.1103/PhysRevLett.121.242301}{\doi{10.1103/PhysRevLett.121.242301}},
\href{http://www.arXiv.org/abs/1801.04895}{\texttt{arXiv:1801.04895}}.

\bibitem{ATLAS:2019dsv}
\hrefCMSnoop {}{{ATLAS Collaboration}, ``{Comparison of fragmentation functions
  for jets dominated by light quarks and gluons from pp and PbPb Collisions in
  ATLAS}'',} \textit{ Phys. Rev. Lett.} \textbf{ 123} (2019) 042001,
  \href{http://dx.doi.org/10.1103/PhysRevLett.123.042001}{\doi{10.1103/PhysRevLett.123.042001}},
  \href{http://www.arXiv.org/abs/1902.10007}{\texttt{arXiv:1902.10007}}.

\bibitem{CMS:2024zjn}
\hrefCMSnoop {}{{CMS Collaboration}, ``{Girth and groomed radius of jets
  recoiling against isolated photons in lead-lead and proton-proton collisions
  at sNN=5.02~TeV}'',} \textit{ Phys. Lett. B} \textbf{ 861} (2025) 139088,
  \href{http://dx.doi.org/10.1016/j.physletb.2024.139088}{\doi{10.1016/j.physletb.2024.139088}},
  \href{http://www.arXiv.org/abs/2405.02737}{\texttt{arXiv:2405.02737}}.

\bibitem{hepdata}
\hrefCMSnoop {}{``{HEPData record for this analysis}'',} 2024.
\newblock
  \href{http://dx.doi.org/10.17182/hepdata.154439}{\doi{10.17182/hepdata.154439}}.

\bibitem{CMS:2008xjf}
\hrefCMSnoop {}{{CMS Collaboration}, ``The {CMS} experiment at the {CERN}
  {LHC}'',} \textit{ JINST} \textbf{ 3} (2008) S08004,
  \href{http://dx.doi.org/10.1088/1748-0221/3/08/S08004}{\doi{10.1088/1748-0221/3/08/S08004}}.

\bibitem{Hayrapetyan_2024}
\hrefCMSnoop {}{{CMS Collaboration}, ``{Development of the CMS detector for the
  CERN LHC Run 3}'',} \textit{ JINST} \textbf{ 19} (2024) P05064,
  \href{http://dx.doi.org/10.1088/1748-0221/19/05/p05064}{\doi{10.1088/1748-0221/19/05/p05064}},
  \href{http://www.arXiv.org/abs/2309.05466}{\texttt{arXiv:2309.05466}}.

\bibitem{CMS-LUM-17-003}
\hrefCMSnoop {}{{CMS Collaboration}, ``Precision luminosity measurement in
  proton-proton collisions at $\sqrt{s} =$ 13 {TeV} in 2015 and 2016 at
  {CMS}'',} \textit{ Eur. Phys. J. C} \textbf{ 81} (2021) 800,
  \href{http://dx.doi.org/10.1140/epjc/s10052-021-09538-2}{\doi{10.1140/epjc/s10052-021-09538-2}},
  \href{http://www.arXiv.org/abs/2104.01927}{\texttt{arXiv:2104.01927}}.

\bibitem{CMS:2025rzq}
\href {https://arxiv.org/abs/2503.03946}{{CMS Collaboration}, ``{Luminosity
  measurement for lead-lead collisions at $\sqrt{s_{\mathrm{NN}}}$ = 5.02 TeV
  in 2015 and 2018 at CMS}'',} 2025.
  \href{http://www.arXiv.org/abs/2503.03946}{\texttt{arXiv:2503.03946}}.
  Submitted to Eur. Phys. J. C.

\bibitem{CMS-PAS-LUM-19-001}
\href {http://cds.cern.ch/record/2765655}{{CMS Collaboration}, ``{Luminosity
  measurement in proton-proton collisions at 5.02 \TeV in 2017 at CMS}'',} {CMS
  Physics Analysis Summary} CMS-PAS-LUM-19-001, 2021.
\newblock {\url{http://cds.cern.ch/record/2765655}}.

\bibitem{Khachatryan:2016bia}
\hrefCMSnoop {}{{CMS Collaboration}, ``{The CMS trigger system}'',} \textit{
  JINST} \textbf{ 12} (2017) P01020,
  \href{http://dx.doi.org/10.1088/1748-0221/12/01/P01020}{\doi{10.1088/1748-0221/12/01/P01020}},
\href{http://www.arXiv.org/abs/1609.02366}{\texttt{arXiv:1609.02366}}.

\bibitem{Khachatryan:2010us}
\hrefCMSnoop {}{{CMS Collaboration}, ``Transverse-momentum and pseudorapidity
  distributions of charged hadrons in pp collisions at {$\sqrt{s} = 7$\TeV}'',}
  \textit{ Phys. Rev. Lett.} \textbf{ 105} (2010) 022002,
  \href{http://dx.doi.org/10.1103/PhysRevLett.105.022002}{\doi{10.1103/PhysRevLett.105.022002}},
  \href{http://www.arXiv.org/abs/1005.3299}{\texttt{arXiv:1005.3299}}.

\bibitem{Sirunyan:2017ulk}
\hrefCMSnoop {}{{CMS Collaboration}, ``{Particle-flow reconstruction and global
  event description with the CMS detector}'',} \textit{ JINST} \textbf{ 12}
  (2017) P10003,
  \href{http://dx.doi.org/10.1088/1748-0221/12/10/P10003}{\doi{10.1088/1748-0221/12/10/P10003}},
\href{http://www.arXiv.org/abs/1706.04965}{\texttt{arXiv:1706.04965}}.

\bibitem{CMS:2020uim}
\hrefCMSnoop {}{{CMS Collaboration}, ``{Electron and photon reconstruction and
  identification with the CMS experiment at the CERN LHC}'',} \textit{ JINST}
  \textbf{ 16} (2021) P05014,
  \href{http://dx.doi.org/10.1088/1748-0221/16/05/P05014}{\doi{10.1088/1748-0221/16/05/P05014}},
  \href{http://www.arXiv.org/abs/2012.06888}{\texttt{arXiv:2012.06888}}.

\bibitem{Cacciari:2011ma}
\hrefCMSnoop {}{M.~Cacciari, G.~P. Salam, and G.~Soyez, ``{FastJet user
  manual}'',} \textit{ Eur. Phys. J. C} \textbf{ 72} (2012) 1896,
  \href{http://dx.doi.org/10.1140/epjc/s10052-012-1896-2}{\doi{10.1140/epjc/s10052-012-1896-2}},
\href{http://www.arXiv.org/abs/1111.6097}{\texttt{arXiv:1111.6097}}.

\bibitem{Berta:2014eza}
\hrefCMSnoop {}{P.~Berta, M.~Spousta, D.~W. Miller, and R.~Leitner,
  ``{Particle-level pileup subtraction for jets and jet shapes}'',} \textit{
  JHEP} \textbf{ 06} (2014) 092,
  \href{http://dx.doi.org/10.1007/JHEP06(2014)092}{\doi{10.1007/JHEP06(2014)092}},
\href{http://www.arXiv.org/abs/1403.3108}{\texttt{arXiv:1403.3108}}.

\bibitem{Khachatryan:2016kdb}
\hrefCMSnoop {}{{CMS Collaboration}, ``{Jet energy scale and resolution in the
  CMS experiment in pp collisions at 8 TeV}'',} \textit{ JINST} \textbf{ 12}
  (2017) P02014,
  \href{http://dx.doi.org/10.1088/1748-0221/12/02/P02014}{\doi{10.1088/1748-0221/12/02/P02014}},
\href{http://www.arXiv.org/abs/1607.03663}{\texttt{arXiv:1607.03663}}.

\bibitem{CMS:2017wyc}
\href {http://cds.cern.ch/record/2256875}{{CMS Collaboration}, ``{CMS} jet
  algorithms performance in 13 {TeV} data'',} CMS Physics Analysis Summary
  CMS-PAS-JME-16-003, 2016.
\newblock {\url{http://cds.cern.ch/record/2256875}}.

\bibitem{Khachatryan:2010fm}
\hrefCMSnoop {}{{CMS Collaboration}, ``{Measurement of the isolated prompt
  photon production cross section in pp collisions at $\sqrt{s} = 7$\TeV}'',}
  \textit{ Phys. Rev. Lett.} \textbf{ 106} (2011) 082001,
  \href{http://dx.doi.org/10.1103/PhysRevLett.106.082001}{\doi{10.1103/PhysRevLett.106.082001}},
\href{http://www.arXiv.org/abs/1012.0799}{\texttt{arXiv:1012.0799}}.

\bibitem{AWES1992130}
T.~C. Awes\hrefCMSnoop {}{ { et~al.}, ``A simple method of shower localization
  and identification in laterally segmented calorimeters'',} \textit{ {Nucl.
  Instrum. Meth. A}} \textbf{ 311} (1992) 130,
  \href{http://dx.doi.org/10.1016/0168-9002(92)90858-2}{\doi{10.1016/0168-9002(92)90858-2}}.

\bibitem{Sjostrand:2014zea}
T.~Sj{\"o}strand\hrefCMSnoop {}{ { et~al.}, ``{An introduction to PYTHIA
  8.2}'',} \textit{ Comput. Phys. Commun.} \textbf{ 191} (2015) 159,
  \href{http://dx.doi.org/10.1016/j.cpc.2015.01.024}{\doi{10.1016/j.cpc.2015.01.024}},
\href{http://www.arXiv.org/abs/1410.3012}{\texttt{arXiv:1410.3012}}.

\bibitem{CMS:2019csb}
\hrefCMSnoop {}{{CMS Collaboration}, ``{Extraction and validation of a new set
  of CMS PYTHIA 8 tunes from underlying-event measurements}'',} \textit{ Eur.
  Phys. J. C} \textbf{ 80} (2020) 4,
  \href{http://dx.doi.org/10.1140/epjc/s10052-019-7499-4}{\doi{10.1140/epjc/s10052-019-7499-4}},
  \href{http://www.arXiv.org/abs/1903.12179}{\texttt{arXiv:1903.12179}}.

\bibitem{Lokhtin:2008xi}
I.~P. Lokhtin\hrefCMSnoop {}{ { et~al.}, ``{Heavy ion event generator HYDJET++
  (HYDrodynamics plus JETs)}'',} \textit{ Comput. Phys. Commun.} \textbf{ 180}
  (2009) 779,
  \href{http://dx.doi.org/10.1016/j.cpc.2008.11.015}{\doi{10.1016/j.cpc.2008.11.015}},
  \href{http://www.arXiv.org/abs/0809.2708}{\texttt{arXiv:0809.2708}}.

\bibitem{CMS:2021vui}
\hrefCMSnoop {}{{CMS Collaboration}, ``{First measurement of large area jet
  transverse momentum spectra in heavy ion collisions}'',} \textit{ JHEP}
  \textbf{ 05} (2021) 284,
  \href{http://dx.doi.org/10.1007/JHEP05(2021)284}{\doi{10.1007/JHEP05(2021)284}},
  \href{http://www.arXiv.org/abs/2102.13080}{\texttt{arXiv:2102.13080}}.

\bibitem{DAgostini:1994fjx}
\hrefCMSnoop {}{G.~D'Agostini, ``{A multidimensional unfolding method based on
  Bayes' theorem}'',} \textit{ Nucl. Instrum. Meth. A} \textbf{ 362} (1995)
  487,
  \href{http://dx.doi.org/10.1016/0168-9002(95)00274-X}{\doi{10.1016/0168-9002(95)00274-X}}.

\bibitem{brenner2020comparison}
L.~Brenner\hrefCMSnoop {}{ { et~al.}, ``{Comparison of unfolding methods using
  RooFitUnfold}'',} \textit{ Int. J. Mod. Phys. A} \textbf{ 35} (2020) 2050145,
  \href{http://dx.doi.org/10.48550/arXiv.1910.14654}{\doi{10.48550/arXiv.1910.14654}},
  \href{http://www.arXiv.org/abs/1910.14654}{\texttt{arXiv:1910.14654}}.

\bibitem{Skwarnicki:1986xj}
\href {https://inspirehep.net/literature/230779}{T.~Skwarnicki, ``{A study of
  the radiative CASCADE transitions between the Upsilon-Prime and Upsilon
  resonances}''}.
\newblock PhD thesis, Cracow, INP, 1986.
\newblock {\url{https://inspirehep.net/literature/230779}}.

\bibitem{Casalderrey-Solana:2012evi}
\hrefCMSnoop {}{J.~Casalderrey-Solana, Y.~Mehtar-Tani, C.~A. Salgado, and
  K.~Tywoniuk, ``{New picture of jet quenching dictated by color coherence}'',}
  \textit{ Phys. Lett. B} \textbf{ 725} (2013) 357,
  \href{http://dx.doi.org/10.1016/j.physletb.2013.07.046}{\doi{10.1016/j.physletb.2013.07.046}},
  \href{http://www.arXiv.org/abs/1210.7765}{\texttt{arXiv:1210.7765}}.

\bibitem{Kunnawalkam_Elayavalli_2017}
\hrefCMSnoop {}{R.~K. Elayavalli and K.~C. Zapp, ``Medium response in {\sc
  jewel} and its impact on jet shape observables in heavy ion collisions'',}
  \textit{ JHEP} \textbf{ 07} (2017) 141,
  \href{http://dx.doi.org/10.1007/jhep07(2017)141}{\doi{10.1007/jhep07(2017)141}},
  \href{http://www.arXiv.org/abs/1707.01539}{\texttt{arXiv:1707.01539}}.

\bibitem{Casalderrey_Solana_2017}
J.~Casalderrey-Solana\hrefCMSnoop {}{ { et~al.}, ``Angular structure of jet
  quenching within a hybrid strong/weak coupling model'',} \textit{ JHEP}
  \textbf{ 03} (2017) 135,
  \href{http://dx.doi.org/10.1007/jhep03(2017)135}{\doi{10.1007/jhep03(2017)135}},
  \href{http://www.arXiv.org/abs/1609.05842}{\texttt{arXiv:1609.05842}}.

\bibitem{Lokhtin_2011}
\hrefCMSnoop {}{I.~P. Lokhtin, A.~V. Belyaev, and A.~M. Snigirev, ``Jet
  quenching pattern at {LHC} in {PYQUEN} model'',} \textit{ Eur. Phys. J. C}
  \textbf{ 71} (2011)
  \href{http://dx.doi.org/10.1140/epjc/s10052-011-1650-1}{\doi{10.1140/epjc/s10052-011-1650-1}},
  \href{http://www.arXiv.org/abs/1103.1853}{\texttt{arXiv:1103.1853}}.

\bibitem{Skands:2014pea}
\hrefCMSnoop {}{P.~Skands, S.~Carrazza, and J.~Rojo, ``Tuning {PYTHIA} 8.1: the
  {Monash} 2013 tune'',} \textit{ Eur. Phys. J. C} \textbf{ 74} (2014) 3024,
  \href{http://dx.doi.org/10.1140/epjc/s10052-014-3024-y}{\doi{10.1140/epjc/s10052-014-3024-y}},
\href{http://www.arXiv.org/abs/1404.5630}{\texttt{arXiv:1404.5630}}.

\end{thebibliography}\endgroup
\cleardoublepage \appendix\section{The CMS Collaboration \label{app:collab}}\begin{sloppypar}\hyphenpenalty=5000\widowpenalty=500\clubpenalty=5000\cmsinstitute{Yerevan Physics Institute, Yerevan, Armenia}
{\tolerance=6000
A.~Hayrapetyan, A.~Tumasyan\cmsAuthorMark{1}\cmsorcid{0009-0000-0684-6742}
\par}
\cmsinstitute{Institut f\"{u}r Hochenergiephysik, Vienna, Austria}
{\tolerance=6000
W.~Adam\cmsorcid{0000-0001-9099-4341}, J.W.~Andrejkovic, L.~Benato\cmsorcid{0000-0001-5135-7489}, T.~Bergauer\cmsorcid{0000-0002-5786-0293}, S.~Chatterjee\cmsorcid{0000-0003-2660-0349}, K.~Damanakis\cmsorcid{0000-0001-5389-2872}, M.~Dragicevic\cmsorcid{0000-0003-1967-6783}, P.S.~Hussain\cmsorcid{0000-0002-4825-5278}, M.~Jeitler\cmsAuthorMark{2}\cmsorcid{0000-0002-5141-9560}, N.~Krammer\cmsorcid{0000-0002-0548-0985}, A.~Li\cmsorcid{0000-0002-4547-116X}, D.~Liko\cmsorcid{0000-0002-3380-473X}, I.~Mikulec\cmsorcid{0000-0003-0385-2746}, J.~Schieck\cmsAuthorMark{2}\cmsorcid{0000-0002-1058-8093}, R.~Sch\"{o}fbeck\cmsAuthorMark{2}\cmsorcid{0000-0002-2332-8784}, D.~Schwarz\cmsorcid{0000-0002-3821-7331}, M.~Sonawane\cmsorcid{0000-0003-0510-7010}, W.~Waltenberger\cmsorcid{0000-0002-6215-7228}, C.-E.~Wulz\cmsAuthorMark{2}\cmsorcid{0000-0001-9226-5812}
% ...
\par}
\vskip\cmsinstskip
\dag:~Deceased\\
$^{1}$Also at Yerevan State University, Yerevan, Armenia\\
$^{2}$Also at TU Wien, Vienna, Austria\\
% ...
\end{sloppypar}
\end{document}